\documentclass[usenatbib,onecolumn]{mn2e}
\usepackage[pass,letterpaper]{geometry}
\usepackage{aecompl}
\usepackage{amssymb}
\usepackage{amsmath}
\usepackage{graphicx}
\usepackage{color}
\usepackage{url}
\usepackage{ifthen}
\usepackage{bm}
\usepackage{hyperref}
\usepackage{aas_macros}
\usepackage{algorithm}
\usepackage{algpseudocode}

\newcommand{\Mpch}{{$h^{-1}$~Mpc}}

\newcommand{\kms}{km s$^{-1}$}
\newcommand{\kmsMpc}{{km s$^{-1}$ Mpc$^{-1}$}}

\newcommand{\mvec}[1]{\bm{#1}}
\newcommand{\mmat}[1]{\mathbf{#1}}
\newcommand{\ScalarAmp}{{A_S}}
\newcommand{\ScalarAmpX}[1]{{A_{S,#1}}}
\newcommand{\virbius}{{\tt VIRBIuS}}
\newcommand{\jcap}{JCAP}

\begin{document}

\bibliographystyle{mn2e}

\title{Bayesian 3d velocity field reconstruction with VIRBIuS}

\author[G. Lavaux]{Guilhem Lavaux$^{1,2}$\thanks{Email: lavaux@iap.fr}\\
$^{1}$ Sorbonne Universit\'es, UPMC Univ Paris 6 et CNRS, UMR 7095, Institut d’Astrophysique de Paris, 98 bis bd Arago,\\ F-75014 Paris, France
}

\date{Accepted XXX. Received XXX; in original form XXX}
\pagerange{\pageref{firstpage}--\pageref{lastpage}}
\maketitle

\label{firstpage}

\begin{abstract}
I describe a new Bayesian based algorithm to infer the full three dimensional velocity field from {observed} distances and spectroscopic galaxy catalogues. In addition to the velocity field itself, the algorithm reconstructs {true} distances, some cosmological parameters and specific non-linearities in the velocity field.  The algorithm takes care of selection effects, miscalibration issues and can be easily extended to handle direct fitting of, e.g., the inverse Tully-Fisher relation. I first describe the algorithm in details alongside its performances. This algorithm is implemented in the \virbius{} (VelocIty Reconstruction using Bayesian Inference Software) software package. I then test it on different mock distance catalogues with a varying complexity of observational issues. The model proved to give robust measurement of velocities for mock catalogues of 3,000 galaxies. I expect the core of the algorithm  to scale to tens of thousands galaxies. It holds the promises of giving a better handle on future large and deep distance surveys for which individual errors on distance would impede velocity field inference. 
\end{abstract}



\section{Introduction}

Peculiar velocities are deviations of the apparent motion of tracers, e.g. galaxies, from the Hubble flow. They are an essential tool to study dynamics of the Local Universe and in particular to probe the underlying gravity field, which is currently assumed to be generated by a Dark Matter density field. At low redshift, i.e. $z \la 0.1-0.2$, they are the only practical way of reconstructing the unbiased, true matter density field. The first mention of galaxy peculiar velocities go back to Hubble \citep{HH31}. When large scale structures data have been first acquired, peculiar velocities have quickly attracted a large attention \citep{AHMS82,ABMHSC86,LFBDDTW88}, before fading out due to a lack of large corpus of distance data and robust methods of analysis.

New distance surveys, from which peculiar velocities can be inferred, have emerged in the recent years like SFI++ \citep{SFIPP06,SFIPP07,SFIPP09}, 6dFv \citep{Campbell14}, CosmicFlows-1 \citep{CosmicFlow1}, CosmicFlows-2  \citep{Tully13}. More surveys are coming online such as TAIPAN/WALLABY \citep{Beutler11,Duffy12}. These surveys revived peculiar velocities as first class probes of cosmology by providing hundreds of thousands of distances. However peculiar velocity analysis is notoriously error prone, being sensitive to different bias and systematic effects, e.g. homogeneous \citep{LFBDDTW88} and inhomogeneous \citep{DBF90} Malmquist bias, distance indicator calibration uncertainties \citep{Willick94} or edge effects. Several attempts have been made at reconstructing the density field directly from distance data. For example one can note the POTENT method \citep{BD89,DBF90,DEKLY99}, the Wiener filter approach \citep{ZHFL95}, or the Unbiased Minimum Variance algorithm \citep{Zaroubi2002}. Additionally, the procedure to derive the power spectrum of the velocity field is relatively complex and prone to the same aforementioned systematics, though there have been some early attempts at measuring it \citep{JK95,KD97,ZZDHK97,Macaulay11,Macaulay12}. New methods have also been recently designed to measure more accurately {the first moments of cosmic flows} from different aspects of distances and luminosities {\citep[e.g.][]{ND11,NBD11,NBD12,FNB14}}.

A common framework capable of handling all  these items at the same time and building a consistent three dimensional (3D) peculiar velocity field is still missing. I am proposing to build such a framework from a full Bayesian joint analysis of the density field (bandwidth restricted Fourier modes of the density field), the cosmological parameters (e.g. $\Omega_\text{m}$, the Hubble constant, the amplitude of scalar fluctuations), the observational parameters (e.g. selection function, Tully-Fisher relation) and the limitations of the model (e.g. amount of small scale non-linearities). By incorporating all these issues in a single framework, this model holds the promise of reducing (maybe cancelling) all systematic effects on the estimation of the 3D peculiar velocity field. The software has been named \virbius{} (VelocIty Reconstruction using Bayesian Inference Software) and will be publicly available later on the author web-page\footnote{\url{http://www.iap.fr/users/lavaux/}.}. 
{Parts of the model will ressemble VELMOD\citep{Willick94,WSDK97}. For example \cite{Willick94} modeled the relation between the true distance and the observables for distance indicators the look like Tully-Fisher relations. Also, \cite{WSDK97} modeled the relation  between redshift observations and Tully-Fisher observables (i.e. magnitudes and HI linewidth). These elements are parts of \virbius{}, but they are generalized and included in a wider framework. I note that \cite{JBK14} have also pushed the effort of measuring accurately velocity field. In all the aforementioned work however, a common framework to handle all components self-consistently are not included. 
Also the possibility of unseen measurement failure is not accounted  for. This will be another major addition (and complexity) to the model. Of course augmenting the model with limited data available comes at a cost: e.g. the powerspectrum must be parameterized in terms of a small number of cosmological parameters. Among them I will select a few of particular interests: the overall amplitude of the powerspectrum, which is degenerate with the growth factor and the Hubble constant, which governs the shape. All the other cosmological parameters are kept fixed in this work. I will introduce other parameters that describe the dataset itself (e.g. zero-point calibration, noise levels). }


The structure of the paper is as follows.
In Section~\ref{sec:stats}, I describe both the adopted model and the algorithm that I have developed to explore the parameter space given some distance galaxy catalogue. The model, in Section~\ref{sec:model}, includes description of cosmological expansion, distance uncertainties, and a clean separation between the linear and the non-linear component of the velocity field. The model is fully Bayesian and priors can be adjusted easily to include more detailed description of selections effects. In Section~\ref{sec:algorithm}, I describe in detail the algorithm that is required to efficiently sample the posterior distribution of all the parameters that enter into the model, including the velocity field itself. In Section~\ref{sec:test_mock}, I present the results of the test of this algorithm on a variety of mock catalogues: an ideal, though slightly unrealistic, and a mock catalogue generated assuming perfect homogeneity of tracers and Gaussian random fields statistics for velocity fields (Section~\ref{sec:test_mock_grf}), a more realistic mock catalogue based on halos of an $N$-body simulation either with a trivial or a more complex selection function (Section~\ref{sec:test_mock_halo}). In Section~\ref{sec:conclusion}, I conclude on the performance of the algorithms and the prospects for its use for existing and future distance surveys.

\section{Statistical Method}
\label{sec:stats}

In this Section, I explain the model that I am using to describe self-consistently the velocity field, the cosmology, the redshifts and the distances of the tracers of the velocity field. In Section~\ref{sec:model}, I detail the model and the approximations that I have made. In Section~\ref{sec:algorithm}, I describe the algorithm used to sample the posterior distribution in the huge parameter space.

\subsection{Model}
\label{sec:model}

\subsubsection{The flow model}

I propose to solve the general problem of reconstructing in an unbiased way the three dimensional peculiar velocity field and cosmological parameters from a set of redshifts and distance modulus of tracers. I put $N_d$ the number of tracers. I propose a self-consistent approach based on a probabilistic modelling. For the low redshift Universe, and a given tracer $i$, it is possible to write a linear relationship between the redshift $z_i$, the distance $d_i$, the pseudo-Hubble constant $\widetilde{H}$ at redshift zero and the peculiar velocity $\mvec{v}(\mvec{r})$ as:
\begin{equation}
	z_i = \widetilde{H} d_i + v^r_i + \epsilon_{z,i},\label{eq:linear_redshift_relation}
\end{equation} 
with $v^r_i = \mvec{v}(d_i \hat{u_i})\hat{\mvec{u}}_i$ the line of sight {component of} the peculiar velocity of the $i$-th object, $\hat{\mvec{u}}_i$ the unit vector pointing in the direction of the tracer, $\epsilon_{z,i}$ the redshift measurement error. This is the usual Hubble relation, though we have replaced $H$ by $\widetilde{H}$ to take into into account the fact that the calibration of distance indicator may not be absolute. Additionally, we do not have access to a precise probe of the distance. The Equation~\eqref{eq:linear_redshift_relation} is only valid at extremely low redshift. The aim of this work is to have a self consistent and accurate reconstruction of velocity field for large and deeper distance survey. Of course different cosmological distances appear, like the luminosity and the comoving distances. From now on we will use $d_i$ as the comoving distance of an object $i$ and $d^\text{L}_i$ its corresponding luminosity distance. The exact relation combining cosmological and peculiar velocity, while they are non-relativistic, induced Doppler effect is the following:
\begin{equation}
	1 + z_i = \left(1 + \bar{z}_i(d^\text{L}_i)\right)\left (1 + \frac{v^r_i}{c}\right), \label{eq:doppler}
\end{equation}
with $\bar{z}_i$ the cosmological redshift, which depends on the luminosity distance, $v^r_i$ the line of sight component of the velocity field, assumed to be small compared to the speed of light $c$ and in the rest frame of large scale structures. The relation, for a flat universe, is explicitly the following { \citep[e.g.][]{Weinberg72}}
\begin{equation}
	\bar{d}^L(\bar{z}) = \frac{c (1+\bar{z})}{\widetilde{H}} \int_{z=0}^{\bar{z}} \text{d}z\; \frac{1}{E(z)},\label{eq:luminosity_distance}
\end{equation}
and
\begin{equation}
	E(z) = \left(\Omega_\text{M} (1+z)^3 + \Omega_\Lambda\right)^{1/2}.
\end{equation}
The comoving distance $\bar{d}(d^L,\bar{z})$ is related to the luminosity distance $d^L$ according to
\begin{equation}
	\bar{d}(d^L, \bar{z}) = \frac{\bar{d}_L}{1+\bar{z}}.
\end{equation}
The Equation~\eqref{eq:luminosity_distance} is numerically invertible which, assuming $d^L$ is known, allows for the derivation of  the cosmological redshift. {In the text, we will introduce the observational counterpart of the luminosity distance, called distance modulus $\mu$, whose definition is:
\begin{equation}
	\mu = 5 \log_{10} \left(\frac{\bar{d}^L}{10~\text{pc}}\right)
\end{equation}}
Finally the relation \eqref{eq:doppler} can be rewritten as followed:
\begin{equation}
	v^r_i = c \frac{z_i - \bar{z}_i(\bar{d}^\text{L}_i)}{1 + \bar{z}_i(d^\text{L}_i)} \label{eq:redshift_relation}.
\end{equation}
\cite{DS14} recently reminded the community that using the linear approximation instead of Equation~\eqref{eq:redshift_relation} leads to substantial error even at relatively low redshift ($z \la 0.05$). So it is fundamental to include the complete treatment in my analysis so that the reconstructed velocity field are unbiased for future peculiar velocity surveys. {I am assuming that the measured redshift is without error in the above equation. Of course that is not the case, and the redshift error will be treated in the next section.}

Though this relation between the $v^r_i$ and the observed redshift is more complex than Equation~\eqref{eq:linear_redshift_relation}, it does not introduce any new {systematic errors}. The only problem that is introduced is the proper tracking of the cosmological redshift $\bar{z}_i$ and the comoving distance $d_i$ when the luminosity distance $d^L_i$ changes. In all this work, all algorithms make use of the Equation~\eqref{eq:redshift_relation} instead of the linear relation~\eqref{eq:linear_redshift_relation}.
{In the above I am considering that the observation of luminosity distance is perfect. That is not the case in practice as a number of effects are changing the apparent luminosity such as gravitational lensing, Integrated Sachs-Wolfe effect and gravitational redshift and peculiar velocities \citep{Sasaki87,PB04,BDG06}. For the moment we will neglect all these effects, keeping in mind that in data they will eventually have to be inserted into the likelihood analysis of luminosity distances. The last of these effects could be important to ensure consistent treatment of peculiar velocities as highlighted by \cite{Sasaki87,HG06}. To summarize the distance modulus itself is affected by peculiar velocities at first order because the observed flux is itself sensitive to beaming and Doppler effects. The observed luminosity distance is in fact \citep{HG06}:
\begin{equation}
	d^L_\mathrm{o} = \bar{d}^L_\mathrm{LSS} \left(1 + \frac{1}{c}(2 \mvec{v}_e - \mvec{v}_o) . \mvec{\hat{n}}\right)
\end{equation}
with $d^L_\mathrm{o}$ the luminosity distance determined in the observer rest frame, i.e. from observed flux, $\bar{d}^L_\mathrm{LSS}$ the actual luminosity distance of the object in an homogeneous universe with a FLRW metric, $\mvec{v}_e$ ($\mvec{v}_o$ respectively) the peculiar velocity of the emitter (observer respectively) with respect to this homogeneous background, and $c$ the speed of light in vacuum. This relation is exact at first order. In this work we will neglect the impact of this term, while focusing on the peculiar velocity present in the Doppler effect of the observed spectrum (Equation~\eqref{eq:doppler}). I note that the introduction of this correction would not change the algorithm fundamentally but introduce additional complexity in the formulation of the likelihood of the distance modulus (as detailed in Equation~\eqref{eq:likelihood}). I also note that most peculiar velocity analysis (except supernovae) neglect the full impact of this term \citep{JBK14}.
}

\subsubsection{The million parameter likelihood analysis}

Historically, direct extrapolation of the velocity field from this relation has lead to a number of biases, like the inhomogeneous and homogeneous Malmquist biases \citep{DBF90}. They originate from the reuse of an imprecise distance indicator for both estimating the line of sight peculiar velocity $v^r_{i} = \mvec{v} . \hat{\mvec{u}}_i$ and {using it as an estimate of the true distance} $q^h d_i$, with $q^h = \widetilde{H}/H$. We are however not doomed to be limited by this problem. I propose to consider the distance itself as a random variable to generate the velocity field. This is not an entirely new proposal. In \cite{WSDK97}, the VELMOD technique was already trying to improve the distance using a likelihood approach. They used a peculiar velocity field predicted using linear theory of gravitational instabilities and galaxy redshift surveys as a prior for the velocity field. The idea of generating random velocity fields in agreement with observation is not new either, it originates back to the constrained realization of Gaussian random fields in cosmological context \citep{HR91,HR92}. There was no published work that has attempted to blend both constrained realizations, distance sampling and parameter estimation. {We are not bound to be limited to proceed sequentially for the analysis of peculiar velocity field. Notably, it is in principle possible to adjust both the power-spectrum and the field itself, as it is done for the data of the Cosmic Microwave Background \citep{WLL04}.}.

Ideally, we would like to have a representation of the joint probability of the velocity field, the distances, the Hubble constant, and possibly cosmological parameters $p$ and an additional noise parameter $\sigma_\text{NL}$, which will characterize the departure from linear theory of the actual velocity field. So in practice we will fit the following model:
\begin{align}
	1 + z_i &= (1 + \bar{z}_i(d^L_i))\left(1 + \frac{\mvec{v}_\text{linear}(q^h d_i \hat{u_i}) . \hat{\mvec{u}}_i  + \epsilon_{\text{NL},i}}{c} \right)  + \epsilon_{z,i}, \\
		&=(1 + \bar{z}_i(d^L_i))\left(1 + \frac{H f \mvec{\Psi}_\text{linear}\left(q^h d_i \hat{u_i}\right) . \hat{\mvec{u}}_i  + \epsilon_{\text{NL},i}}{c} \right) + \epsilon_{z,i}, \label{eq:model}
\end{align} 
with
\begin{equation}
	\langle \epsilon_{\text{NL},i} \epsilon_{\text{NL},j} \rangle = \sigma_{\text{NL},type(i)}^2 \delta_{i,j}. \label{eq:variance_nl}
\end{equation}
$d_i$ the comoving distance from the observer of the tracer $i$ (actually a function of $d^L_i$ in this work),
and $H$ the Hubble constant at redshift $z=0$.  Each tracer is given an assignment $type(i)$. { I note that $d_i$ is the comoving coordinates scaled with $\tilde{H}$. So it is not the true distance but the distance in the convention of the calibration of the distance indicator. If the distance ladder is correctly built then $q^h=1$, however this is not in general guaranteed. In Equation~\eqref{eq:model} we have introduced the displacement field $\mvec{\Psi}$, which is expressed in comoving coordinates.  If we assume that no vorticity is created on large scales, it can be simply described through its divergence, which I will call $\Theta$ to follow earlier conventions on velocity fields.  In the Lagrangian linear regime, the displacement field is related to the velocity field by a linear relation, which gives the second part of Equation~\eqref{eq:model}. Thus, we have the following relations:
\begin{align}
	\mvec{\nabla}_{\mvec{r}}.\mvec{\Psi} & = \Theta(\mvec{r}), \\
	\mvec{v}_\text{linear}(\mvec{r}) & = f H \mvec{\Psi}_\text{linear}(\mvec{r}).
\end{align}
Physically, $\Theta$ is related to the density fluctuations at present time in the Universe owing to continuity equation \citep[e.g][]{Peebles80}. Additionally I will call $\hat{\Theta}=\{\hat{\Theta}(\mvec{k}_q)\}$  
the discrete Fourier basis which represent the full continuous field $\Theta$. Thus their formal relationship is:
\begin{equation}
	\Theta(\mvec{r}) = \frac{1}{L^3} \sum_{\mvec{q}} \hat{\Theta}(\mvec{k}_q)  \mathrm{e}^{\text{i}  \mvec{k}_{\mvec{q}} . \mvec{r}},
\end{equation}
with
\begin{equation}
   \mvec{k}_q = \frac{2\pi}{L} \mvec{q}
\end{equation}
and $\mvec{q} \in \{0,1,\ldots,N-1\}^3$, $N$ the resolution of the reconstructed field.
These amplitudes will serve as free parameters of the field in the rest of the text. The splitting of the velocity into the scaling and displacement components allows us to make a shortcut later on when adjusting the Hubble constant to the data.}
The most simple model (Equation~\ref{eq:variance_nl}) has a single type, but it is possible to have several. This approach can be required if we try to model a set of tracers which could be split into sub-population such as clustered and non-clustered (i.e. elliptical vs spiral galaxies). {It presents the other advantage of isolating potential catastrophic errors in the distance or spectroscopic measurement of a tracer. For example, the assignment of a supernova to a galaxy is sometimes dubious as its observed spectrum can be heavily blue shifted by the explosion processes. In other cases, the adjusted luminosity distance can be a strong outlier in statistical empirical relation such as the Tully-Fisher relation.} {The introduction of several types of tracers (including outliers) is related to the problem of the } Gaussian mixture \citep{Pearson1894,DempsterLairdRubin77}. I evaluate the velocity field at the real distance using the scaling factor $q^h=\widetilde{H}/H$. 
$\mvec{v}_\text{linear}(\mvec{r})$
has the same statistical properties as the velocity field derived from linear perturbation theory at $z=0$.  

{Finally the data are given by duet for each galaxy $i$: the distance modulus $\mu_i$ and the observed redshift $z_i$. I will assume that the noise on the observed distance modulus and the redshift measurement are both Gaussian, with standard deviations $\sigma_{\mu,i}$ and $\sigma_{z,i}$ respectively. Because the two data are acquired independently} the likelihood, i.e. the probability of observing the data $\{ (\mu_i, z_i) \}$, given the model is immediately given by:
\begin{multline}
	\mathcal{L} = P(\{\mu_i, z_i \} | \{ d^L_i \}, \{ \sigma_{z,i}, \sigma_{\mu,i} \}, \{ \hat{\Theta}(\mvec{k}_q) \}, H, \widetilde{H}, \Sigma_{\text{NL}}, \mathcal{T}, \{ p^{type}_t \} ) \\
	   \propto \prod_{i=1}^{N_d} \left(\sigma_{z,i}^2 (1+\bar{z}_i)^{-2} + \sigma_{\text{NL},type(i)}^2 \right)^{-1/2} \times\\ \exp\left\{-\frac{1}{2}\sum_{i=1}^{N_d} \frac{\left[v_i^r(z_i,d_i)-H f \Psi_{r,i}(q^h)\right]^2}{ \left[(\sigma_{z,i}^2 (1+\bar{z}_i(d_i))^{-2} + \sigma_{\text{NL},type(i)}^2\right]} -   
    \frac{(\mu_i - 5\log_{10}(d^L_i/10\text{ pc}))^2}{\sigma_{\mu,i}^2}\right\}, \label{eq:likelihood}
\end{multline}
with $\Psi_{r,i}(q^h) = \mvec{\Psi}_\text{linear}(q^h d_i \mvec{\hat{u}}_i) . \hat{\mvec{u}}_i$, $\Sigma_\text{NL}=\{\sigma_{\text{NL},q}\}$, $\mathcal{T} = \{ type(q) \}$ {and $N_d$ the number of provided tracers (i.e. the size of set $\{\sigma_{\mu,i}\}$).}
Using Bayes identity, we may now express the posterior probability of the parameters, given the data:
{\begin{multline}
	P(\mathcal{D}^L=\{ d^L_i \}, \hat{\Theta}=\{ \hat{\Theta}(\mvec{k}_q) \}, \widetilde{H},  H, \ScalarAmp, \{\sigma_{\text{NL},t}\}, \mathcal{T} | \\	
	  \mathcal{M}=\{\mu_i\}, \mathcal{Z}=\{z_i\}, \Sigma_z = \{\sigma_{z,i}\}, \Sigma_\mu=\{\sigma_{\mu,i} \}) = \\
\frac{%
\mathcal{L} \times \pi(\mathcal{D}^L) \pi(\hat{\Theta})  \pi(\Sigma_{\text{NL}}) \pi(H)  \pi(\{type(i)\}) \pi(\ScalarAmp)}%
{\sum_{\mathcal{T}'}\int\text{d}H \text{d}\hat{\Theta} \text{d}\mathcal{D}^L \text{d}\Sigma_\text{NL}\; \mathcal{L} \times \pi(\mathcal{D}^L) \pi(\mathcal{T}') \pi(\hat{\Theta}) \pi(\{\sigma_{\text{NL},t}\}) \pi(H)}, \label{eq:main_posterior}
\end{multline}
where $\mathcal{T}'$ runs over all possible type combinations in the denominator, $t$ is the index of one of the type.}
The functions $\pi$ are priors on the specified parameters. 
I assume that the statistics of $\hat{\Theta}$ is {determined} by the power spectrum $P(k)$ of primordial density fluctuations scaled to redshift zero. This power-spectrum may itself depends on some cosmological parameters. I have chosen to incorporate the Hubble constant $H$ and the amplitude of the power spectrum $\ScalarAmp$ as free parameters but to keep the other parameters at their best-fit value from other probes, as e.g. WMAP9 \citep{WMAP9_COSMO} or Planck \citep{PLANCK_COSMO}. {$\ScalarAmp$ is defined as the pre-factor in the unnormalized power spectrum:
\begin{equation}
	P_{\Theta\Theta}(k) = \ScalarAmp \left(\frac{k}{1 h\text{ Mpc}^{-1}}\right)^{n_\text{s}} T^2(k),
\end{equation}
with $T(k)$ the transfer function, normalized to one for $k\rightarrow 0$.
}
These free parameters will allow us to run a self-consistent check of the cosmology. We thus have:{
\begin{equation}
	\pi(\hat{\Theta}|H,\ScalarAmp) = \prod_{\mvec{q}} (2\pi P_{\Theta\Theta}(k_{\mvec{q}}; H,\ScalarAmp))^{-1/2}\exp\left(-\frac{|\hat{\Theta}(\mvec{k}_{\mvec{q}})|^2}{2 P_{\Theta\Theta}(k_{\mvec{q}};H,\ScalarAmp)}\right). \label{eq:prior_theta}
\end{equation}}
The natural basis of representation of $\hat{\Theta}$ is thus the Fourier basis. 
{Because $\mvec{\Psi}$ is assumed to be without vorticity, it can be solved with the same Green function as the gravity. 
We can introduce an auxiliary scalar field $\Phi$ such that 
$\mvec{\Psi} = \mvec{\nabla}_{\mvec{x}} \Phi$ and $\Phi$ must satisfy the Poisson equation $\Delta \Phi = \Theta$. 
Thus in Fourier space we obtain:
\begin{equation}
	\Phi(\mvec{r}) = - \sum_{\mvec{q}} \frac{1}{k_{\mvec{q}}^2} \text{e}^{i \mvec{k}_{\mvec{q}}.\mvec{r}} \hat{\Theta}(\mvec{k}_{\mvec{q}}) \label{eq:phi_fourier}
\end{equation}}
 The full expression of the displacement $\mvec{\Psi}$ in terms of $\hat{\Theta}$ is thus {after taking the gradient of $\Phi$:}
\begin{equation}
	\mvec{\Psi}(\mvec{r}) = \sum_{\mvec{q}} \frac{i \mvec{k}_{\mvec{q}}}{k_{\mvec{q}}^2} \text{e}^{i \mvec{k}_{\mvec{q}}.\mvec{r}} \hat{\Theta}(\mvec{k}_{\mvec{q}}) \label{eq:psi_fourier}
\end{equation}

\subsubsection{Some notes on priors}

The prior on the distance translates our preconception of the localization of the tracers in the volume into a mathematical {expression}. I chose to consider two possibilities. The {first} kind of prior that I consider consists in fitting an empirical distribution $f_g(d)$ of galaxies assuming isotropy and uniformity in the choice of the tracers. The distribution is self-consistently estimated from the ensemble of reconstructed distances, and thus in unit of Mpc. The empirical distribution is given by
\begin{equation}
	\pi_\text{isotropy}(\mathcal{D}^L|p,d_\text{cut},n) \propto D^p \exp\left(-\left(\frac{D}{d_\text{cut}}\right)^n\right),\label{eq:selfunction}
\end{equation}
which is trivially combined in the total distance prior
\begin{equation}
	\pi_\text{isotropy}(\mathcal{D}^L|p,d_\text{cut},n) \propto \prod_{i=1}^{N_d} f_g(d^{L}_{i}; p, d_\text{cut}, n).
\end{equation}
Note that the use of this prior expands the parameter space to include $\{ p,d_\text{cut},n \}$, and {we use the luminosity distance of the tracers.}

{For some mock catalogue, I will consider a second choice, i.e. that tracers are homogeneously distributed in a given determined by luminosity distance. While this choice is questionable as the tracers are more expected to be homogeneously distributed in comoving volume, this choice simplifies greatly the tests. It also does not remove any value to the \virbius{} model as in any practical case the first prior will be used, which automatically absorb differences between luminosity distances and comoving distances in the parameterization. Thus the prior takes the form:
\begin{equation}
	\pi_\text{homogeneous}(\mathcal{D}^L) \propto \prod_{i=1}^{N_d} (d^L_i)^2.
\end{equation}
Finally, for the very Local Universe, the comoving and luminosity distances are equal, thus this prior correspond to a classical problem.
In particular, this prior is related to the "homogeneous Malmquist bias" correction \citep{LB88,StraussWillick95}, which leaves the inhomogeneous part not modeled. \cite{StraussWillick95} indicates that the correction introduced by the inhomogeneous component is sub-dominant in their simulation. Of course this statement depends on the statistical distribution of the tracers themselves, as elliptical galaxies will be located more in the center of the density peaks. We will put the homogeneous approximation of the prior to the test in Section~\ref{sec:test_mock}. }

Each tracer $i$ is given a type $type(i)$. The prior probability of this typing is given by a finite set of values:
\begin{equation}
	\pi(\mathcal{T}|\mathcal{P}) = \prod_{i=1}^{N_d} p^{type}_{type(i)},
\end{equation}
where $\mathcal{P}$ is the ensemble of possible probabilities $\{ p^{type}_j \}$ with $j$ going over the available types, {and $type(i)$ maps the $i$-th galaxy to the $j$-th type}.

Finally, I assume a uniform prior on the Hubble constant $H$, the effective Hubble constant $\widetilde{H}$ and on the variances $\sigma^2_{\text{NL},q}$. I acknowledge that the uniform prior on $\widetilde{H}$ is not equivalent to a uniform prior on the zero-point calibration of the distance indicator, which are often linearly derived from magnitudes. Assuming a uniform prior on the magnitude of the zero-point would yield a prior $\pi(\widetilde{H}) \propto 1 / \widetilde{H}$ which is stricter than a pure uniform prior on $\widetilde{H}$. 

In the above model, I have not treated the problem of selection effects. There could be some concern that the selection function of catalogues is not modelled here. Indeed according to \cite{Willick94,StraussWillick95}, depending on the choice of the used distance indicator (e.g. forward Tully-Fisher vs inverse Tully-Fisher), some systematic bias could be introduced in the velocity/distance reconstruction. I will argue in the following section that they have nearly  no effect on the algorithm except in the determination of distances. 

\subsection{Sampling algorithms}
\label{sec:algorithm}

I use the blocked Gibbs sampling method \citep{LIU01031994,GG84,WLL04} to solve for the problem of having an unbiased estimate of the velocity field, including proper error bars on all parameters of the adopted model. This sampling technique is related to Markov Chains such as the Metropolis-Hasting algorithm \citep{MU49,MRRTT53,H70}, but in this case we always accept the new proposed move. Blocked Gibbs sampling is converging efficiently in two cases: cosmic variance limited problems and high signal to noise ratio regimes. Unfortunately, it has potentially long convergence when model parameters are correlated and/or {the model has to face} intermediate signal-to-noise ratio regimes. Gibbs sampling has the advantage of splitting a complicated posterior into pieces that are easier to compute. I note that we have a Markov chain whose state $\mathcal{M}_i$ is described by the vector \footnote{The last three parameters are only involved for the more detailed selection function.}
\begin{align}
	\mathcal{M}_i &= \big( H_i; \widetilde{H}_i; \ScalarAmpX{i};  \Sigma_{\text{NL},i}; \hat{\Theta}_i; \mathcal{D}^L_i; \mathcal{P}_i; \mathcal{T}_i; d_{\text{cut},i}; p_i, n_i), \\
	 &= \big( s_{i,j} \big) \label{eq:markov_order}
\end{align}
with $\hat{\Theta}$ the Fourier modes of the velocity field, sampled on a finite grid of modes $\mvec{k}_q$, $\mathcal{T}_i$ the typing of tracers (noted $\{ type(q) \}$ above) and $\mathcal{P}_i$ the probability of each type. I remind the reader that $n$, $p$ and $d_\text{cut}$ are the parameters of the model for the selection function given in Equation~\eqref{eq:selfunction}. I note $N_S$ the number of variables in $\mathcal{M}_i$. The second equation (Equation~\ref{eq:markov_order}) implicitly defines the ordering of the parameters in the state $\mathcal{M}_i$. I will use the following notation to indicate that I will condition on everything except the  indicated variable $s_{i,j}$: 
\begin{align}
	\bar{s}_{i,j} &= \big(s_{i,0},\ldots,s_{i,j-1},s_{i-1,j+1},\ldots,s_{i-1,N_S} \big).
\end{align}
However if I have analytically marginalized according to the velocity potential $\hat{\Theta}$ then I note:
\begin{align}
	\widetilde{\bar{s}}_{i,j} &= \bar{s}_{i,j} \backslash (\hat{\Theta}_i)
\end{align}
In general the algorithm will produce a new chain state using the Algorithm~\ref{alg:markov}.
\begin{algorithm}
\begin{algorithmic}[1]
  \Procedure{generateMarkovChainElement}{s}
  \For{\texttt{$j=0$} to $N_S$}
    \State $s_{i,j} \curvearrowleft P(s_{\_,j} | {\bar{s}}_{i,j})$
  \EndFor
  \EndProcedure
  \State
  \Procedure{generateMarkovChain}{}
  	\State $s \leftarrow 0$
    \Loop
    	\State \Call{generateMarkovChainElement}{$s$}
        \State write state $s$ in a file
    \EndLoop
  \EndProcedure
\end{algorithmic}
\caption{\label{alg:markov} Blocked Sampling algorithm  }
\end{algorithm}
with $s_{\_, j}$ indicating that we do not consider the specific value of the parameter $s_j$ but the general posterior of this parameter.

The above algorithm corresponds to the canonical Gibbs-Sampling algorithm. However, some parameters are strongly correlated to $\hat{\Theta}$, as e.g. $\widetilde{H}$.  To improve the convergence, I will use the partially collapsed Gibbs Sampler algorithm \citep{vanDykPark2008}. In the context of this work, it is possible to analytically marginalize according to $\hat{\Theta}$ when $s_{\_,j}$ is $H$, $\widetilde{H}$, $\sigma_{NL,q}$ and $A_S$. This adds complexity to each of the conditional posterior but it is still numerically tractable because the concerned posteriors are mono-dimensional. Defining:
\begin{align}
	\mathcal{M}^0_i & = \left(H_i; \widetilde{H}_i; \ScalarAmpX{i}; \Sigma_{\text{NL},i}\right) = \left(s^0_{i,j}\right), \text{ and }\\
	\mathcal{M}^1_i & = (\mathcal{D}^L_i; \mathcal{P}_i;\mathcal{T}_i;d_{\text{cut},i};p_i,n_i) = \left(s^1_{i,j}\right),
\end{align}
the new algorithm {is given in Algorithm~\ref{alg:partial_collapse} in which we used } $N^q_s$ to specify the number of elements in $\mathcal{M}^q_i$. The probability $P_c$ is obtained by analytically marginalizing the conditional posterior $P(s_{\_,j}, \hat{\Theta} | \widetilde{\bar{s}}_{i,j})$ according to $\hat{\Theta}$.
Each of the used conditional posterior can be deduced from the main posterior \eqref{eq:main_posterior}. I will now detail them one by one. I give in Table~\ref{tab:explicit_conditional_parameters} the parameters onto which each conditional probability function depends explicitly. The last element of the sampling chain corresponds to the sampling of the parameters of the distance selection prior, which in the program is done slightly separately and generates the three parameters of the prior in a single call. 

\begin{algorithm}
	\begin{algorithmic}[1]
	  \Procedure{generatePartiallyCollapsedMarkovChainElement}{s}
	  \For{\texttt{$j=0$} to $N^0_S$}
	    \State $s_{i,j} \curvearrowleft P_c(s^0_{\_,j} | {\widetilde{\bar{s}}^0}_{i,j}, {\widetilde{\bar{s}}^1}_{i,j})$
	  \EndFor
	  \State Generate a constrained realization $\hat{\Theta}_i$
	  \For{\texttt{$j=0$} to $N^1_S$}
	    \State $s_{i,j} \curvearrowleft P(s^1_{\_,j} | {\bar{s}^0}_{i,j}, \hat{\Theta}_i, {\bar{s}^1}_{i,j})$
	  \EndFor
	  \EndProcedure
	\end{algorithmic}
	\caption{\label{alg:partial_collapse} Partially collapsed Gibbs Sampling algorithm  in \virbius. }
\end{algorithm}

Before detailing each of the algorithm required to sample the parameters, I note that this approach allows for alleviating potential systematic biases arising from selection.
The formalism that I gave in the previous section can be transformed to allow for a more complete fitting procedure that includes the distance indicator itself. The incorporation of the effects of selection in the distance relation fitting procedure consists then in simply multiplying the likelihood by the selection function $S(m,\eta,d)$ in the notations of \cite{StraussWillick95}, $m$ the apparent magnitude, $\eta$ the luminosity linewidth, $d$ the distance. As the algorithm uses conditional posteriors to explore the parameter space, this selection function will disappear from all expressions except the one that concerns the distance. If the selection is separable between $(m,\eta)$ and $d$, the use of the forward Tully-Fisher algorithm would even not be sensitive to any details of the selection in $(m,\eta)$. Of course these assumptions are relatively strict and it may be that the parameter space is more entangled, e.g. between $m$ and $d$, which would not allow this simplification. {This remark is related to the distance indicator based on Tully-Fisher, though other indicators could benefit albeit on a different set of variables for the selection function. The entire algorithm presented in this paper is nevertheless completely general.}

I now review each of the conditional posterior one by one to derive their expression.

\begin{table}
	\begin{center}
		\begin{tabular}{c|c|c}
			\hline
			Parameter & Parameter name $s_{\_,j}$ & Explicit dependency \\
			\hline
			\hline
			Physical Hubble constant & $H$ & $\mathcal{D}^L, \widetilde{H}, {\Sigma}_{\text{NL}},\ScalarAmp$ \\
			Distance zero point calibration & $\widetilde{H}$ & $\mathcal{D}^L, H, {\Sigma}_{\text{NL}}$  \\
			Non-linear/spurious error model & ${\Sigma}_{\text{NL}}$ & $\hat{\Theta}, \mathcal{P}, \ScalarAmp$ \\
			Amplitude of scalar fluctuations & $\ScalarAmp$ & $\mathcal{D}^L, H, \widetilde{H}, {\Sigma}_{\text{NL}}$ \\
			\hline
			Velocity field scalar mode & $\hat{\Theta}$ & $\mathcal{D}^L, \Sigma_{\text{NL}}, H, \widetilde{H},\ScalarAmp$ \\
			\hline
			Luminosity distances & $\mathcal{D}^L$ & $\hat{\Theta}, \widetilde{H}, H$ \\
			Type probability & $\mathcal{P}$ & $\mathcal{T}$ \\
			Type & $\mathcal{T}$ & $\mathcal{D}^L, \mathcal{P}, \Sigma_\text{NL}$ \\
			\hline
			Distance prior effective distance & $d_\text{cut}$ & $\mathcal{D}^L$ \\
			Distance prior slope parameter & $p$ & $\mathcal{D}^L$ \\
			\hline
		\end{tabular}
	\end{center}
	\caption{\label{tab:explicit_conditional_parameters} Dictionary for the notation of the parameters used  in this work and explicit parameters onto which the conditional probability explicitly depends.}
\end{table}

\subsubsection{The Hubble constant $H$}
\label{sec:cond_H}

The conditional posterior of the Hubble constant may be derived from the main posterior expression \eqref{eq:main_posterior}. After marginalization according to $\hat{\Theta}$, it gives:\footnote{I am not explicitly giving the dependency of all the terms in this expression, which would be too notation heavy. I am keeping all dependencies implicit. For example, the covariance matrix $[\mmat{C}^w]$ actually depends on $H$, $\widetilde{H}$, $\sigma_\text{NL}$, $\ScalarAmp$ and all distances.}
\begin{equation}
  P(H | \mathcal{D}^L, \mathcal{Z}, \Sigma_z, \widetilde{H}, \ScalarAmp, \Sigma_\text{NL}, \mathcal{T}) \propto |2\pi [\mmat{C}^w(H)]|^{-1}
     \exp\left(-\frac{1}{2}\sum_{i,j=1}^{N_d} v^r_i v^r_j [\mmat{C}^w(H)]^{-1}_{i,j}\right) \label{eq:conditional_H}
\end{equation}
with $v^r_i$ as given by Equation~\eqref{eq:redshift_relation}, the residual velocity once the apparent Hubble flow is subtracted. We have also used the covariance matrix $C^w_{i,j}$, which, following the model of Equation~\eqref{eq:model}, is defined as
\begin{align}
	[\mmat{C}^w]_{i,j} &= \langle v^r_i v^r_j \rangle \\
	&= 
	  \langle \epsilon_{i,NL}^2 \rangle + \langle \epsilon_{i,z}^2 \rangle + 
	     \sum_{a,b=1}^3 {\hat{u}}^{(i)}_{a}  {\hat{u}}^{(j)}_{b} \langle v_{\text{linear},a}(d_i  q^h \mvec{\hat{u}}^{(i)})  v_{\text{linear},b}(d_j q^h \mvec{\hat{u}}^{(j)}) \rangle \\
	       &= \left(\sigma_\text{NL,type(i)}^2 + \sigma^2_{z,i} (1+\bar{z}_i)^{-2}\right) \delta_{i,j}  + C^{v,r}_{i,j} \label{eq:covar_vr_field}
\end{align}
where $\hat{u}^{(i)}_{a}$ refers to the $a$-th component of the unit vector pointing in the direction of the $i$-th galaxy,
\begin{equation}
	C^{v,r}_{i,j} = \langle v_{r,i} v_{r,j} \rangle =  \sum_{\mu,\nu=1}^{3} \hat{u}^{(i)}_\mu \hat{u}^{(j)}_\nu [C^{v}_{\mu,\nu}]_{i,j} \label{eq:covar_vfield},
\end{equation}
$q^h=\widetilde{H} / H$,
and
\begin{equation}
	[\mmat{C}^{v}_{\mu,\nu}]_{,i,j} =  (f H)^2  \int\frac{\text{d}^3\mvec{k}}{(2\pi)^3}\; \frac{k_\mu k_\nu}{|\mvec{k}|^4} \text{e}^{i q^h \mvec{k}.(d_i \mvec{\hat{u}}^{(i)} - d_j\mvec{\hat{u}}^{(j)})} P_{\Theta\Theta}(|\mvec{k}|)
\end{equation}
is the covariance matrix of the large scale part of the velocity field. The covariance matrix $\mmat{C}^v_{\mu,\nu}$ is derived from the prior on $\hat{\Theta}$ given in Equation~\eqref{eq:prior_theta}. {In practice the matrix $\mmat{C}^{v} $ is computed using the Fourier-Taylor algorithm of Appendix~\ref{app:interpolation}.}
The conditional probability given in Equation~\eqref{eq:conditional_H} is mixing the need of small residual and correlated fluctuations through cosmic flows. It may be highly non-Gaussian and not necessarily with a single maximum. It is however a one-dimensional posterior, which makes it possible to tabulate it. I have used the algorithm of Appendix~\ref{app:1d_sample} to generate a random sample from this density distribution.

\subsubsection{The effective Hubble constant $\widetilde{H}$}

I have separated the Hubble constant presently linked to auto-correlations of the velocity field from the one corresponding to the redshift-distance relation. In Section~\ref{sec:cond_H}, we have obtained the (complicated) posterior of the $H$. 
The conditional posterior of $\widetilde{H}$, again marginalized according $\hat{\Theta}$ takes the same form as \eqref{eq:conditional_H}, except that $\widetilde{H}$ is left free and $H$ is kept constant. As $\widetilde{H}$ is involved in more non-linear relations due to cosmological redshift effects, the conditional posterior is non-Gaussian in several aspects. The full expression of this probability density is 
\begin{equation}
	  P(\widetilde{H} | \mathcal{D}^L, \mathcal{Z}, \Sigma_z, H, \ScalarAmp, \Sigma_\text{NL}, \mathcal{T}) \propto |2\pi [\mmat{C}^w(\widetilde{H})]|^{-1}
     \exp\left(-\frac{1}{2}\sum_{i,j=1}^{N_d} v^r_i(\widetilde{H}) v^r_j(\widetilde{H}) [\mmat{C}^w(\widetilde{H})]^{-1}_{i,j}\right) \label{eq:conditional_Htilde},
\end{equation}
with $\mmat{C}^w$ defined as in the previous section.

\subsubsection{The tracer types}
\label{sec:tracer_type_fit}

The model includes some freedom on the type of non-linearity (modelled by the extra noise $\sigma_{\text{NL},k}$ as determined in Section~\ref{sec:sigmaNL_fit}) that affects each tracer. The model that I have adopted is the Gaussian mixture where each type is given an unconditional probability and the adopted extra noise depends on the type. The typing mechanism is represented by the projection function $type(k)$. In the Gibbs sampling framework we can assume that we know the value of $\{ \sigma_{\text{NL},a} \}$ and infer statistically the unconditional probability of the type $type(k)$. The probability the object $k$ has a type $type(k)$ equal to $q$ is thus proportional to the probability of the type multiplied by the probability that the error term is likely according to the numbers derived in Section~\ref{sec:sigmaNL_fit}. 
Mathematically, the likelihood (equivalently the probability) that the set of tracers $\{k\}$ has some  residuals $\{ \epsilon_k \}$ given the type projector $\mathcal{T}$ and the probabilities of typing $\mathcal{P}$ is
\begin{equation}
	P(\{ \epsilon_k \} | \mathcal{T}, \mathcal{P} ) \propto \prod_{k=1}^{N_t} p^{type}_{type(k)} \frac{1}{\sqrt{\sigma^2_{\text{NL},type(k)} + \sigma^2_{z,k} (1+\bar{z}_k)^{-2}}}\exp\left(-\frac{\epsilon_k^2}{2 (\sigma^2_{\text{NL},type(k)} + \sigma^2_{z,k} (1+\bar{z}_k)^{-2})}\right)  \label{eq:type_probability}
\end{equation}
with $\epsilon_a =  v_a^r(z_a, d^L_a) - H f \Psi^r_{a}(q^h)$.
Using Bayes identity, we can derive the conditional probability of the type mapper $\mathcal{T}$ given the residuals $\{\epsilon_k \}$ and the type probability $\mathcal{P}$:
\begin{equation}
	P(\mathcal{T} = \mathcal{T}^r | \{ \epsilon_k \},   \mathcal{P}) = \frac{P(\{ \epsilon_k \} | \mathcal{T}^r, \mathcal{P} ) \pi(\mathcal{T}^r)}{\sum_{\mathcal{T}'} P(\{ \epsilon_k \} | \mathcal{T}', \mathcal{P} ) \pi(\mathcal{T}') } 
	=  \prod_{k=1}^{N_t} P(type(k)=q | \Sigma_\text{NL}, p^{type}_q, \epsilon_k).
\end{equation}
with, for a uniform prior on the type:
\begin{equation}
	P(type(k)=q | \Sigma_\text{NL},\{p^{type}_{q'}\}_{\{q'\}}, \epsilon_k) = 
	\frac{p^{type}_{q} (\sigma^2_{\text{NL},q} + \sigma^2_{z,k}(1+\bar{z})^{-2})^{-1/2}\exp\left(-\frac{\epsilon_k^2}{2 (\sigma^2_{\text{NL},q} + \sigma^2_{z,k} (1+\bar{z}_k)^{-2}) }\right)}
	{\sum_{q'} p^{type}_{q'} (\sigma^2_{\text{NL},q'} + \sigma^2_{z,k} (1+\bar{z})^{-2})^{-1/2} \exp\left(-\frac{\epsilon_k^2}{2 (\sigma^2_{\text{NL},q'} + \sigma^2_{z,k} (1+\bar{z}_k)^{-2}) }\right)}. \label{eq:cond_type_proba}
\end{equation}

To sample the adequate type for the tracer $k$, a brute force approach is largely sufficient: for each tracer we compute the $N_t$ probabilities given by Equation~\eqref{eq:cond_type_proba}, generate a random number $r \in [0,1[$ and choose the type $b$ that satisfies
\begin{equation}
	b =\max \left\{ c 	\left| \sum_{q=1}^c P(type(k) = q | \Sigma_{\text{NL}}, \{p^{type}_{q'}\}_{\{q'\}}, \epsilon_k) \le r \right. \right\}.
\end{equation}
This type $b$ is then assigned to the tracer $a$ for the state $i$ of the Markov Chain.

\subsubsection{The tracer probability}
\label{sec:tracer_probability_fit}

For this step, we assume that the type of each tracer is known and we want to infer the probability of unconditionally typing a particle to the type $q$.
This can be readily derived from Equation~\eqref{eq:type_probability} using Bayes identity:
\begin{equation}
	P(\mathcal{P} | \mathcal{T}) = \frac{\prod_{q=1}^{N^t} p_q^{|T_q|}}{\int_{\mathcal{P}'} \prod_{q=1}^{N^t} {p'}_q^{|T_q|} \mathrm{d}\mathcal{P}'},
\end{equation}
with $\mathcal{P}'$ the set of all possible probabilities such that
\begin{align}
	\sum_{q=1}^{N^t} p_q &= 1, \\
	0 < p_q < 1 & \;\text{for all}\; 1 \le q \le N^t.
\end{align}
with $T_q = \{ k | type(k) = q\}$, and $|T_q|$ the number of elements in $T_q$. This probability density is 
a Dirichlet distribution. I am using the function \verb+gsl_ran_dirichlet+ of the GNU Scientific Library (GSL) to generate samples of such a distribution, conditioned on the number of tracers $|T_q|$ in each type $q$.

\subsubsection{Model error $\sigma_{\text{NL},k}$}
\label{sec:sigmaNL_fit}

We consider here the amount of extra noise $\sigma_\text{NL}$ that is not captured by the part of the model that uses linear perturbation theory to derive the velocity field. The conditional posterior distribution, as derived from the main likelihood \eqref{eq:main_posterior}, is:
\begin{equation}
	P(\sigma^2_{\text{NL},k} | \hat{\Theta}, \widetilde{H}, H, \mathcal{D}^L, \mathcal{Z}, \Sigma_z) \propto 
	  \prod_{i / type(i)=k} \left(\sigma_{z,i}^2 (1+\bar{z}_i)^{-2} + \sigma_{\text{NL},k}^2\right)^{-1/2} \times 
	   \exp\left(-\frac{\epsilon_i^2}{2(\sigma_{z,i}^2 (1+\bar{z}_i)^{-2}  + \sigma_{\text{NL},k}^2)}\right).
\end{equation}
The posterior is written in terms of $\sigma^2_{\text{NL},k}$, as I have indicated that I am taking a uniform prior on $\sigma^2_{\text{NL},k}$ and not $\sigma_{\text{NL},k}$ (Section~\ref{sec:model}). This is again a one-dimensional distribution and I use the algorithm of Appendix~\ref{app:1d_sample}.

\subsubsection{Power normalization $\ScalarAmp$}

The normalization of the power spectrum of $\Theta$ is left free in the model. This allows to account for the possibility of a different growth rate of perturbations or a different amplitude of scalar perturbations of our Local Universe. Again, we are faced with a mono-dimensional conditional posterior distribution. The problem shares a lot of similarities with Section~\ref{sec:cond_H}. We change parameter and write {$\ScalarAmp=\alpha \ScalarAmpX{i-1}$}. The conditional posterior, marginalized according to $\hat{\Theta}$ becomes
\begin{equation}
  P(\alpha |\mathcal{D}^L, \mathcal{Z}, \Sigma_z, \widetilde{H}, H, \sigma_\text{NL}) \propto |\alpha \mmat{C}^{v,r} + \mmat{N}_\text{NL}|^{-1/2} 
     \exp\left(-\frac{1}{2} \sum_{i,j=1}^{N_d} v^r_i(z_i,\bar{z}_i)[\alpha \mmat{C}^{v,r} +\mmat{N}_\text{NL}]^{-1}_{i,j} v^r_j(z_j,\bar{z}_j)\right).
\end{equation}
with $[\mmat{N}_\text{NL}]_{i,j} = \left(\sigma^2_{\text{NL},type(i)} + \sigma^2_{z,i} (1+\bar{z}_i)^{-2}\right) \delta_{i,j}$.
In the above equation, we have re-used the covariance matrix $\mmat{C}^{v,r}$ of the line-of-sight component of the velocity fields sampled at the tracer positions. This matrix is given in Equation~\eqref{eq:covar_vfield}.

\subsubsection{The scalar potential of the velocity field}

Now we consider the conditional probability of the Fourier modes of $\hat{\Theta}$, the opposite of the divergence of the velocity field. After simplification, the conditional posterior takes the following form:
\begin{equation}
	P( \hat{\Theta} | \mathcal{Z}, \mathcal{D}^L, H, \Sigma_{\text{NL}} ) 
	\propto 
	  \exp\left(-\sum_{i=1}^{N_d} \frac{(v^r_i(z_i,\bar{z}_i) - H f \Psi_{r,i})^2}{2(\sigma_{z,i}^2 (1+\bar{z}_i)^{-2} + \sigma_{\text{NL},type(i)}^2)}\right) \times 
	    \exp\left(-\sum_{q=1}^{N_q} \frac{|\hat{\Theta}(\mvec{k}_q)|^2}{2 P_{\Theta\Theta}(|\mvec{k}_q|)}\right).
\end{equation}
Sampling from this probability consists in generating a Gaussian random field satisfying $P(k)$, but with some integrals constrained by redshift observations. As the errors on constraints are Gaussian, we may use the algorithm proposed by \cite{HR91,HR92}. For this I remind the reader of the algorithm. The constrained random field $\hat{\Theta}^\text{CR}$ is built from a random part $\hat{\Theta}^\text{RR}$  and the correlated part:
\begin{equation}
	\hat{\Theta}^{CR}(\mvec{k}) = \hat{\Theta}^{RR}(\mvec{k}) + \langle \hat{\Theta}(\mvec{k}) c_i \rangle C^{w,-1}_{ij} (c_i - \widetilde{c}^\text{RR}_i), \label{eq:hoffmanrybak}
\end{equation}
with $c_i$ the $i$-th constraint to apply, $\widetilde{c}^\text{RR}_i$ the mock observation of the same constraint in the the pure random realization $\hat{\Theta}^{RR}(\mvec{k})$, $C^w_{ij} = \langle c_i c_j \rangle$, as given in Equation~\eqref{eq:covar_vr_field}, is the covariance matrix of the constraints. By construction, {in the infinite volume limit,} we have 
\begin{equation}
	\langle \hat{\Theta}^{RR}(\mvec{k}) \hat{\Theta}^{RR}(\mvec{q}) \rangle = (2\pi)^3 \delta_D(\mvec{k}+\mvec{q}) P_{\Theta\Theta}(k),
\end{equation}
with $\delta_D$ the Dirac distribution. 
In the case of this work, the contraints $c_j$ are line-of-sight component of the velocity field. For the $j$-th tracer, the constraint $c_j$ is:
\begin{equation}
	c_j = \sum_{\mu=1}^3 \hat{\mvec{r}}_{j,\mu} \int \frac{\text{d}^3\mvec{k}}{(2\pi)^3}\; \frac{i k_\mu}{k^2} \text{e}^{i d_j \mvec{k}.\hat{\mvec{r}}_j} F_\text{NL}(\mvec{k}) \hat{\Theta}(\mvec{k}) + \epsilon_{j,\text{NL}} + \epsilon_{j,z},
\end{equation}
where $\hat{r}_{j,\mu}$ is the $\mu$-th component of $\hat{r}_j$ the sky direction of the $j$-th tracer, $\epsilon_{j,\text{NL}}$ ($\epsilon_{j,z}$ respectively) is corresponding to the non-linear component not captured by $\hat{\Theta}$ (the redshift measurement error respectively). I added a filter $F_\text{NL}(\mvec{k})$ to remove the contribution of modes that are below the scale of non-linearity. Effectively it is a Heaviside function on the norm of $\mvec{k}$:
\begin{equation}
	F_\text{NL}(\mvec{k}) = \left\{\begin{array}{rl}
	    1 & \text{if } |\mvec{k}| < k_\text{NL}, \\
	    0 & \text{otherwise}.
	 \end{array}\right.
\end{equation}
The correlation between $\hat{\Theta}$ and $c_j$ is
\begin{equation}
	\langle \hat{\Theta}(\mvec{q}) c_j \rangle = 
		\sum_{\mu=1}^3 \hat{r}_{j,\mu} \int \frac{\text{d}^3\mvec{k}}{(2\pi)^3}\; \frac{i k_\mu}{k^2} \text{e}^{i d_j \mvec{k}.\hat{\mvec{r}}_j} \langle \hat{\Theta}(\mvec{k}) \hat{\Theta}(\mvec{q})\rangle = 
		-\frac{i \hat{\mvec{r}}_{j}.\mvec{q}}{q^2} \text{e}^{- i d_j \mvec{q}.\hat{\mvec{r}}_j} P_{\Theta\Theta}(|\mvec{q}|). \label{eq:c_grf_wiener}
\end{equation}
The equation above is computed globally using all tracers with the algorithm in Appendix~\ref{app:ftaylor_wiener}. If we did not use this algorithm we would have needed $\mathcal{O}(N_d \times N_g)$, with $N_g$ the number of grid elements $\mvec{q}$. On the other hand, the Fourier-Taylor Wiener algorithm allows the same value to be computed in $\mathcal{O}(N_d) + \mathcal{O}(N_g)$.
As indicated in Section~\ref{sec:cond_H}, the covariance matrix is obtained by applying the Fourier-Taylor algorithm of Appendix~\ref{app:interpolation} from the Fast Fourier Transform (FFT) of the weighed power spectrum on a regular grid. I do not use the analytically exact expression  \citep[as given in e.g.][]{Gorski88} because it leads to neglecting finite grid effects and periodic boundary effects which are dominant for the reasonable physical grid sizes which I consider (with typical a side length of 500~Mpc). We compute the mock observations $\widetilde{c}_j$ on the unconstrained field  $\hat{\Theta}^{RR}$ in exactly the same way: we do the Fast Fourier Transform on a grid and then do a Fourier-Taylor synthesis to obtain the velocity field values. Finally we can add a random realization of the noise to the interpolated value to construct $\widetilde{c}_j$. All values are recombined using Equation~\eqref{eq:hoffmanrybak} to obtain the final constrained Gaussian random field.

\subsubsection{Galaxy distances}

The conditional posterior of the distances may be derived from the main posterior expression \eqref{eq:main_posterior} as
\begin{equation}
	P(\mathcal{D}^L | \mathcal{Z}, \Sigma_z, \mathcal{M}, \Sigma_\mu, H, \sigma_\text{NL}) \propto \mathcal{L} \times \pi(\mathcal{D}^L)
   \propto \prod_{i=1}^{N_d} p^d_i(d^L_i),
\end{equation} 
with
\begin{multline}
	p^d_i(d^L) =  \frac{1}{\sqrt{2\pi (\sigma_\text{NL}^2 + \sigma_{z,i}^2 (1+\bar{z}_i)^{-2})}} \times \\
	     \exp\left( -\frac{1}{2(\sigma_\text{NL}^2 + \sigma_{z,i}^2 (1+\bar{z}_i)^{-2})} \left(v^r_i(z_i,\bar{z}_i(d^L)) - f H \Psi_r(d(d^L) \hat{\mvec{u}}_i)\right)^2\right) \times \\
	    \exp\left(\frac{(\mu_i - 5\log_{10}(d^L/10\text{ pc}))^2}{2\sigma_{\mu,i}^2}\right).
\end{multline}
I note that the conditional posterior distribution of the distances of each tracers is separable in $N_d$ independent mono-dimensional conditional posterior. This comes from the assumption that the noise on the redshift measurement is uncorrelated from tracer to tracer and that all correlations between tracer is accounted for by the velocity field and, possibly, the bias function. Selection function and clustering bias would typically be retained in this expression. Both can be added multiplicatively to $p^d_i$. For example it can be $S(m_i,\eta_i,d)$ for the selection function and $(1+ b \delta(d\hat{r}_i))$ for the clustering bias {as already stated for VELMOD class of models \citep{Willick94,StraussWillick95}}. The problem of sampling from this posterior reduces here to a sampling problem from $N^d$ mono-dimensional posteriors. To achieve that, I use the classical algorithm of computing the inverse of the cumulative distribution applied on a random realization of a uniform distribution bounded by $[0,1[$. The displacement field is computed at the appropriate position using the Fourier-Taylor synthesis algorithm of Appendix~\ref{app:interpolation}.

\subsubsection{The distance prior parameters}

As the selection function of the galaxies in distance catalogues is poorly known we have introduced in Section~\ref{sec:model} a flexible selection function that depends on three parameters $\{p, d_\text{cut}, n \}$. The conditional likelihood of the distances is exactly equal to the prior \eqref{eq:selfunction} in this case. Using Bayes identity and a uniform prior on the aforementioned parameters, the sampling of this parameters is achieved by another block sampling step. The probability of one of the three parameters, e.g. $p$, given the others is, using Bayes identity,
\begin{equation}
	P(p | d_\text{cut}, n, \mathcal{D}^L) = \frac{P(p,\ldots)}{\int \text{d}p\; P(p,\ldots)} = \frac{\pi(\mathcal{D}^L | p, d_\text{cut}, n)}{\int\text{d}p\; \pi(\mathcal{D}^L | p, d_\text{cut}, n)},
\end{equation}
{where $P(p,\ldots)$ is the posterior probability of Equation~\eqref{eq:main_posterior}, with explicit dependency on the parameters of the prior $\pi(\mathcal{D}^L)$. All functions simplifies except the explicit dependence linking the distances to the distance selection, despite the possible existence of a selection function on galaxy properties (like the apparent/absolute magnitude or the HI linewidth for Tully-Fisher relation).}
Such a sampling step is achieved using the algorithm of Appendix~\ref{app:1d_sample}. To ensure a proper decorrelation, we loop over this block a number of times (typically ten).


\begin{figure}
	\begin{center}
		\includegraphics[width=.8\hsize]{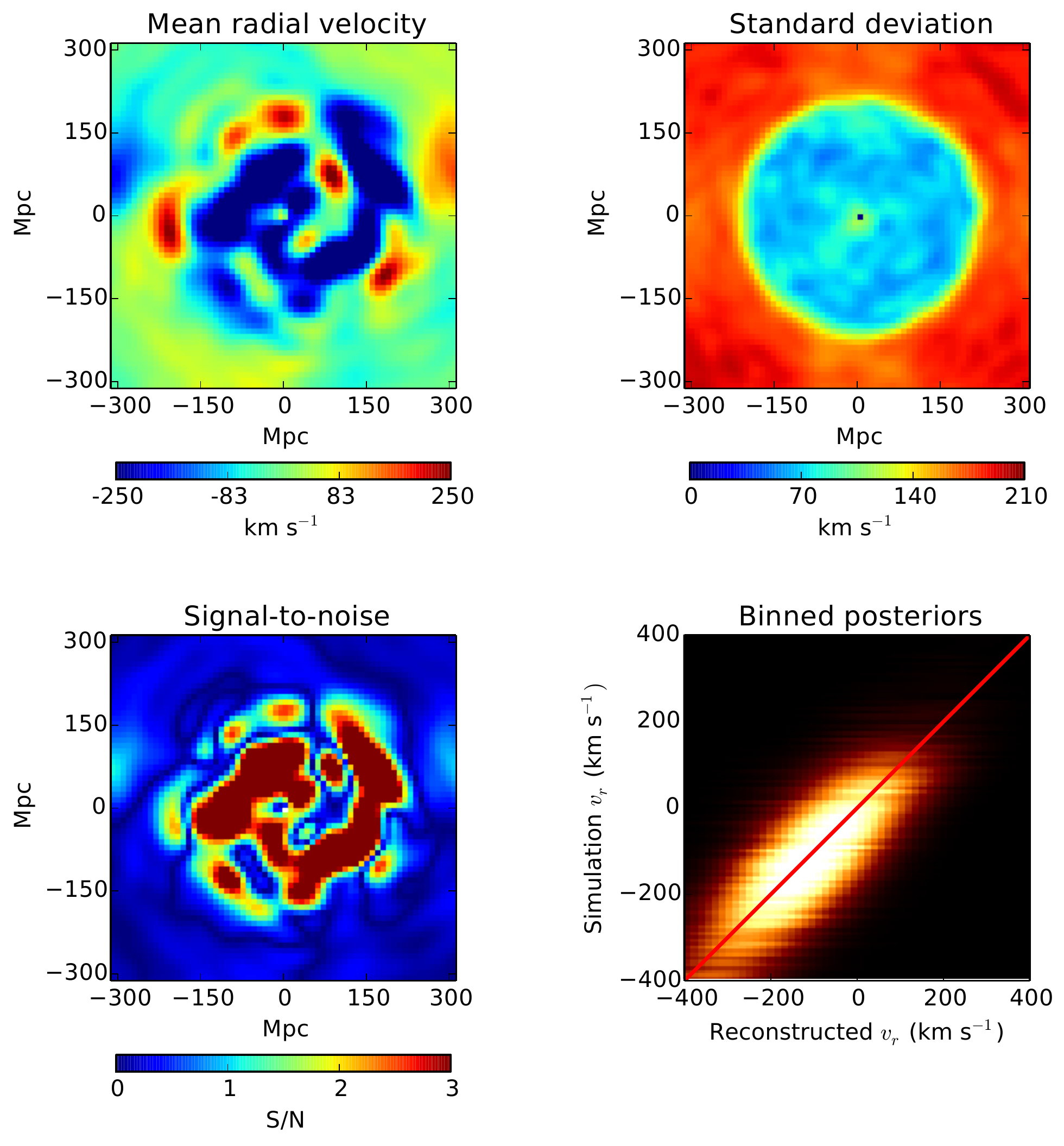}
   \end{center}
   \caption{\label{fig:grf_precise_vr}  Results of the test on the mock catalogue based on Gaussian random field: {central slice of the } ensemble averages for the line of sight of the velocity field (top-left panel), the variance expressed as a standard deviation (top-right panel), the resulting signal to noise (bottom left panel). {In the bottom right panel, I present} a Bayesian comparison between the reconstructed velocities and the actual true velocity field of the simulation using the entire set of posterior distributions.   The details are given in Section~\ref{sec:test_mock_grf}. Only the voxels whose centres are within 150~Mpc from the observer are considered in this panel. {In all panels the selection is isotropic, thus the absence of note on the axis.}}
\end{figure}

\begin{figure}
	\begin{center}
		\includegraphics[width=.8\hsize]{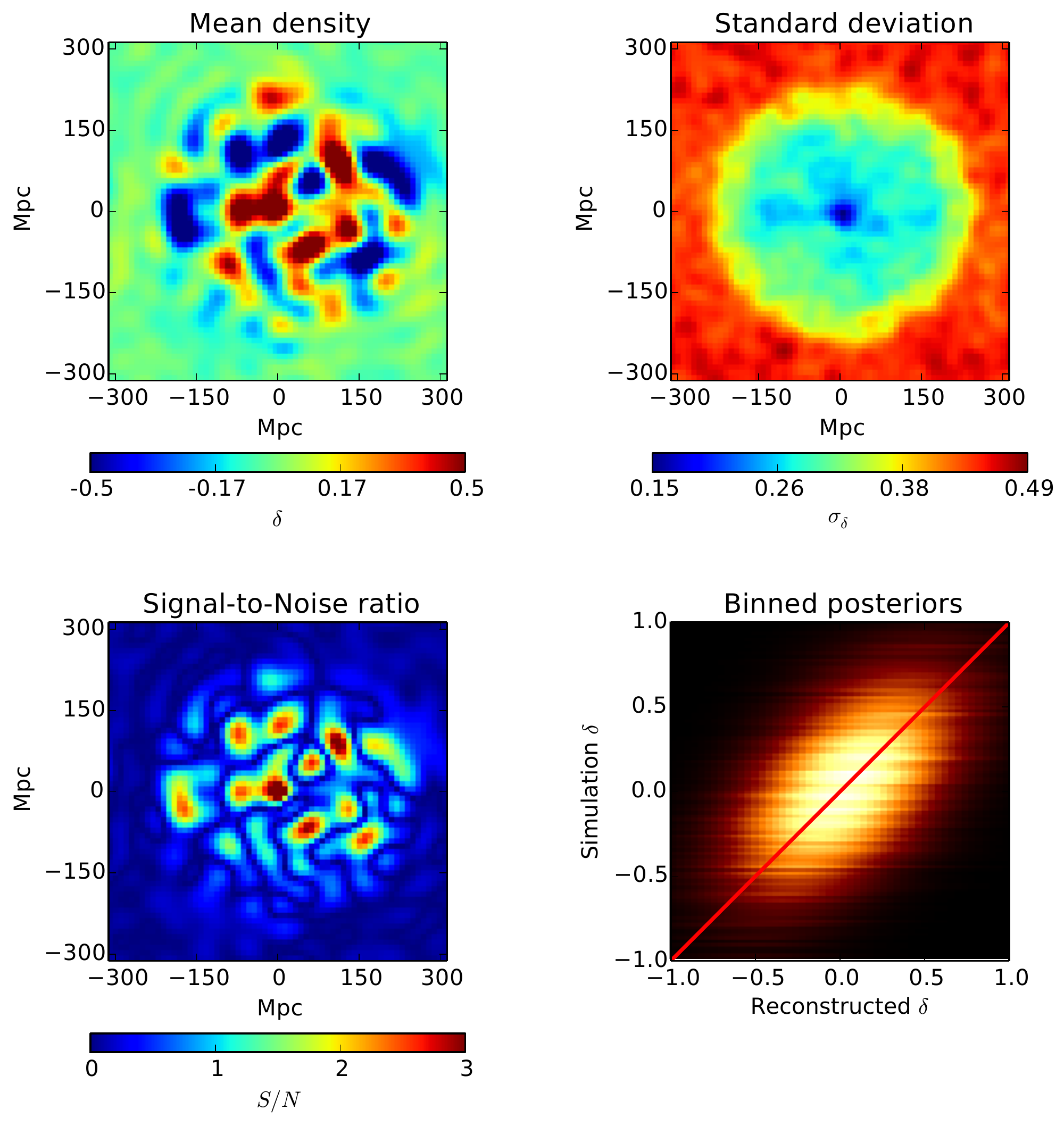}
   \end{center}
   \caption{\label{fig:grf_precise_density} Results of the test on the mock catalogue based on Gaussian random field, density field distribution:  {central slice of the}  ensemble average (top-left panel), the variance expressed as a standard deviation (top-right panel), the resulting signal to noise (bottom left panel). {In the bottom right panel, I present} a Bayesian comparison between the reconstructed densities and the actual true density  field of the simulation using the entire set of posterior distributions. {In all panels the selection is isotropic, thus the absence of note on the axis.}}	
\end{figure}

\begin{figure}
	\begin{center}
		\includegraphics[width=.49\hsize]{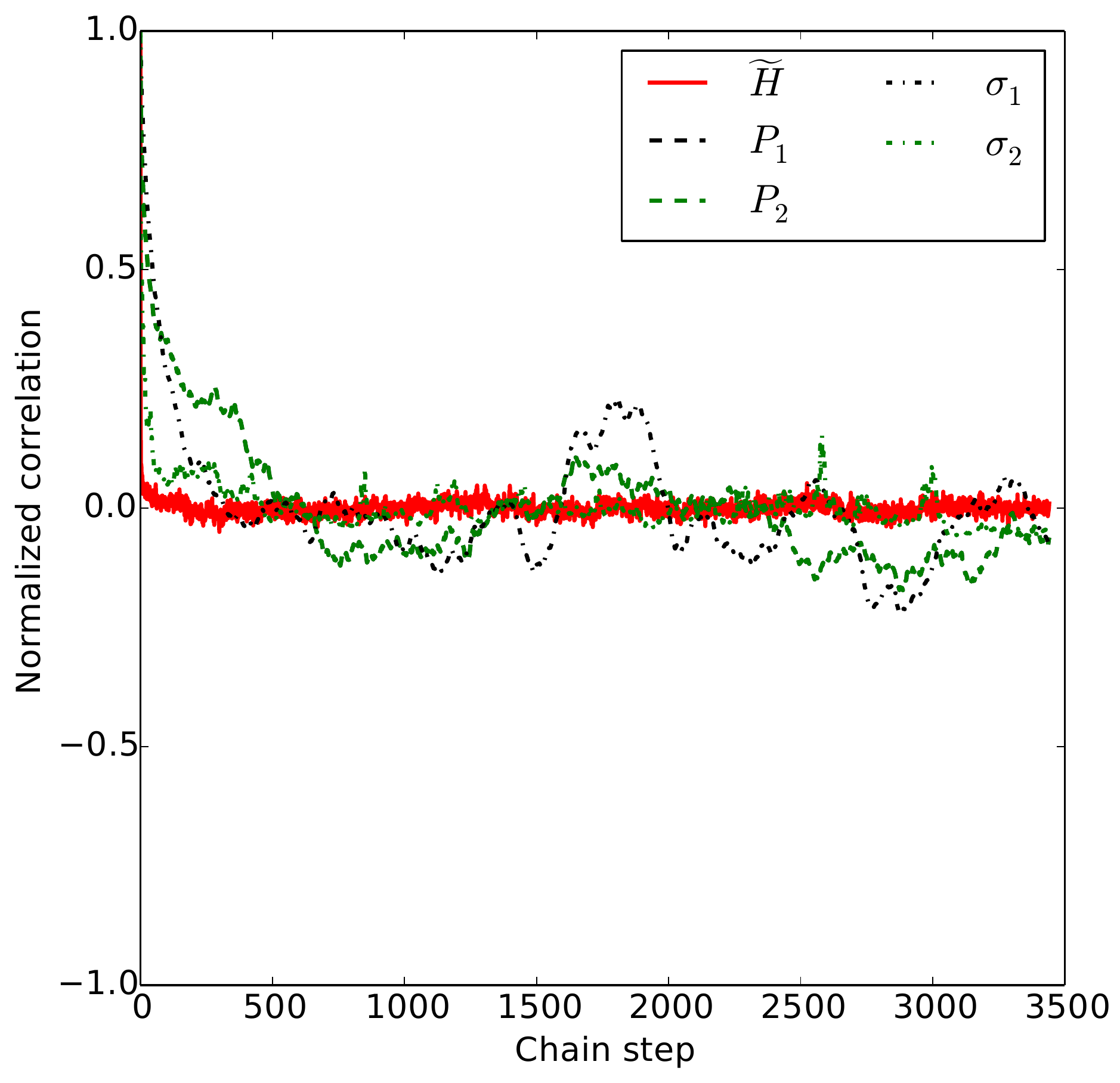}
		\includegraphics[width=.49\hsize]{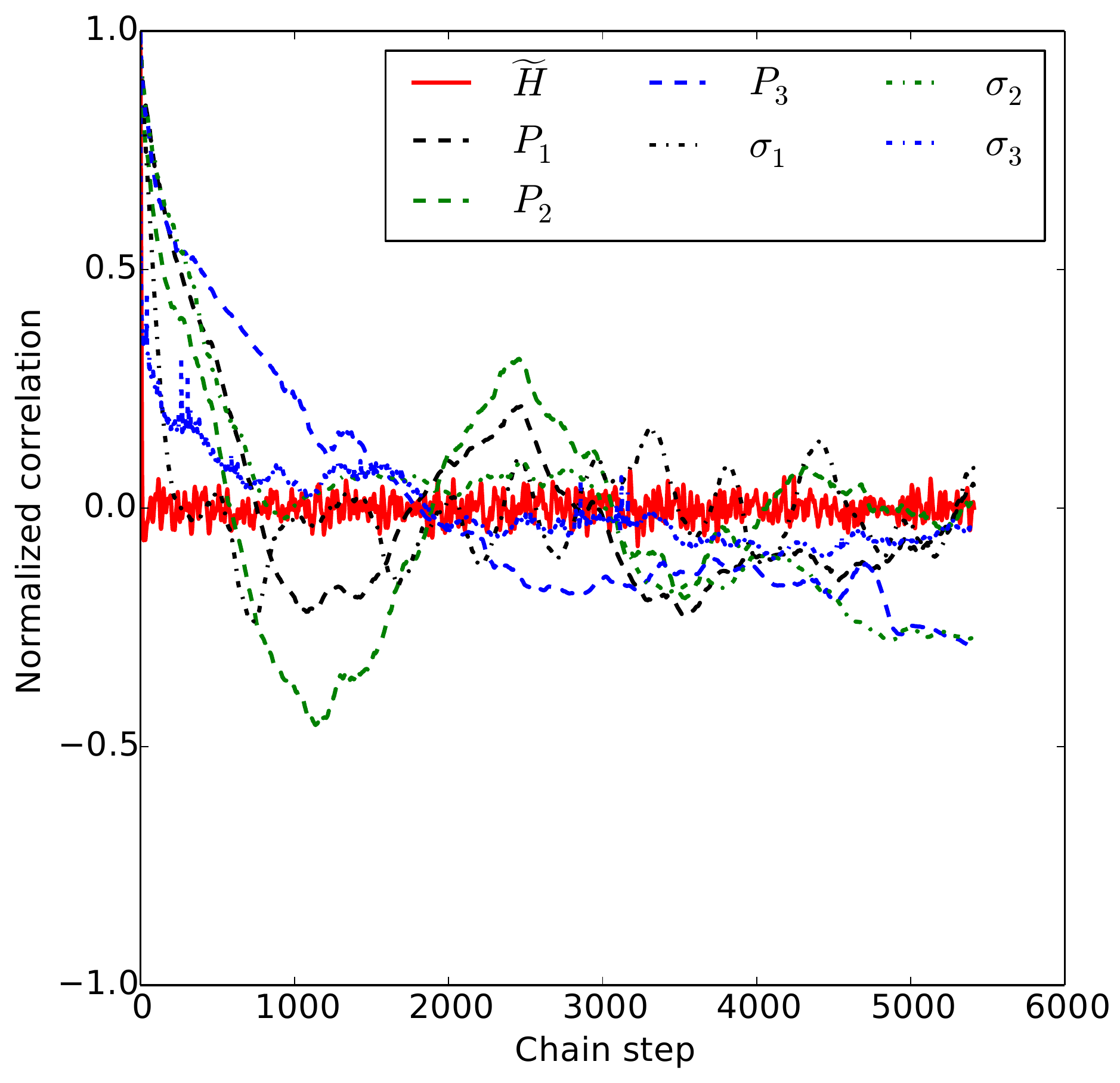}
		\includegraphics[width=.49\hsize]{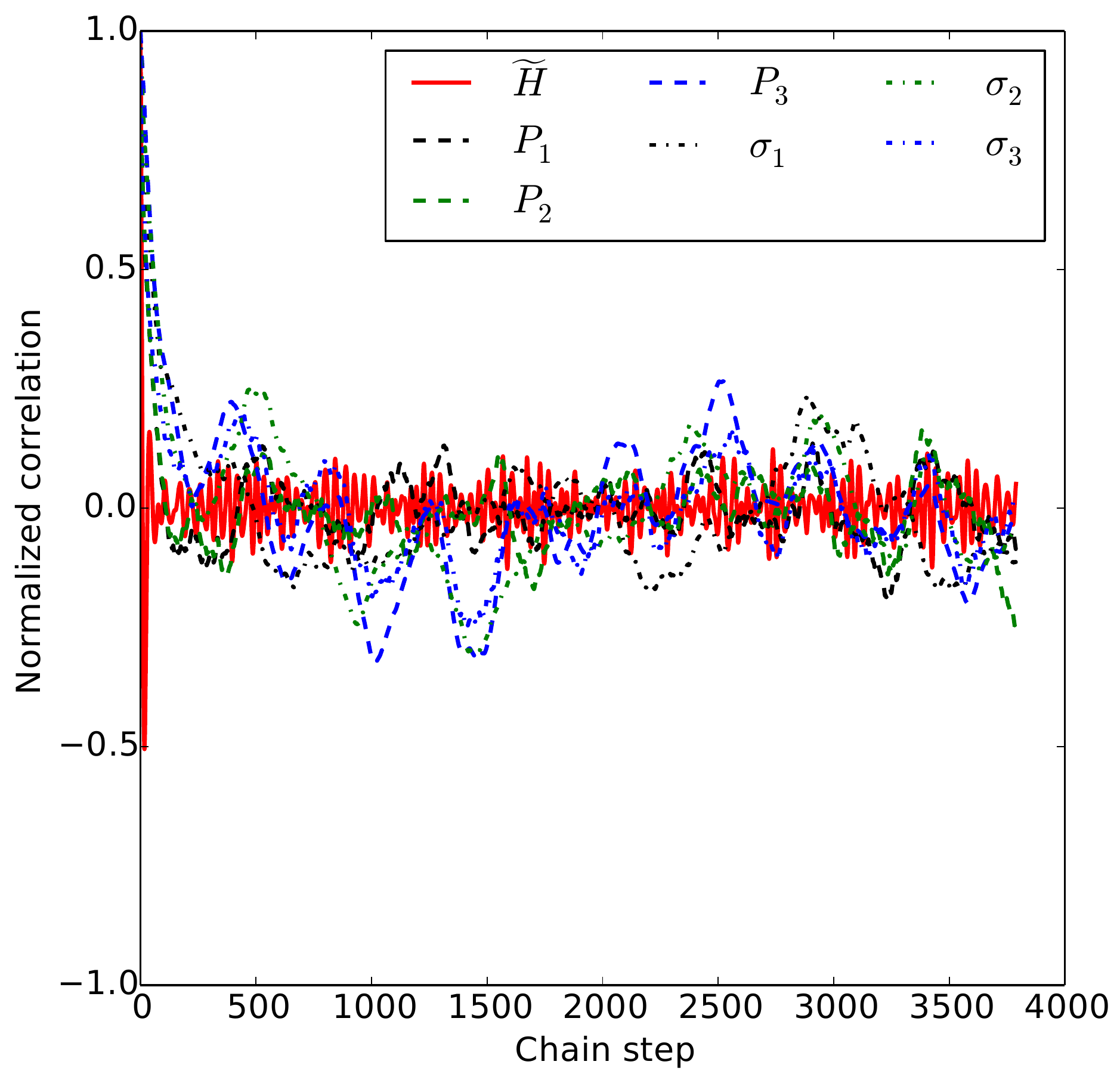}
	\end{center}
	\caption{\label{fig:autocorr_chain} Autocorrelation of the value taken by meta-parameters for the three chains considered in this work: $\widetilde{H}$ (thick solid red), the mixture probability  $P_i$ ($i \in \{1,2,3\}$), dashed lines, respectively in black, green and blue, $\sigma_i$ ($i \in \{1,2,3\}$), the error standard deviation for each tracer type, dash-dotted lines, respectively in black, green and blue. The different panels correspond respectively to the homogeneous Gaussian random field case (top-left), the H3000 catalog (top-right), the Hcomplex catalog (bottom middle). The horizontal axis tracks  the step number along each of the chain.}
\end{figure}

\section{Tests on mock catalogs}
\label{sec:test_mock}

The method that I have described in Section~\ref{sec:stats} is relatively complex and involves two major component: the model of the peculiar velocity field as traced by the galaxies or the clusters of the galaxies and the algorithm itself to adjust the data to the model. This involves two separate sets of tests. First, I will focus on test of the algorithm itself and the performance of the fit on data that were produced to correspond exactly to the model. This is the objective of the Section~\ref{sec:test_mock_grf}. Second, I will look into the capabilities and the limits of the model at reconstructing the velocity field and distances from noisy more realistic mock data sets, typically halo catalogues from $N$-body simulation, which is the objective of Section~\ref{sec:test_mock_halo}.

\subsection{Gaussian random field based mock catalogues}
\label{sec:test_mock_grf}

In this Section, I present the generation and the results of the test of the code against idealized mock tracer catalogues. These catalogues are generated in such a way that the statistics and properties of the tracers follow exactly the model presented in Section~\ref{sec:model}. However they are slightly unrealistic by removing  aspects not captured by the model, such as non-linearities or correlations between velocity field and tracer positions. These aspects may lead to biases in the results. It is nonetheless an interesting exercise to evaluate the performance of the algorithm.

I generate the mock catalogue of tracers, which should be galaxies, as follows:
\begin{enumerate}
	\item I generate a random realization of a "density field", with power spectrum given by linear theory linearly extrapolated to $z=0$ using the expression of \cite{EH98} for the power spectrum of density fluctuations, without the wiggles. The power spectrum is truncated at $k_\text{max}=0.1$~Mpc$^{-1}$.
	\item I pick 3~000 randomly located tracers within a sphere of 200~Mpc, assuming that the tracers are homogeneous in  luminosity distance coordinates. The luminosity distance is then converted in comoving distance and cosmological redshift.
	\item I compute the velocity field at a resolution of $N=512$, and evaluate its value at the position of tracers using a trilinear interpolation.
	\item I compute the line-of-sight component {of the peculiar velocity} and the distance modulus. The mock observables are then generated taking a homogeneous noise of $\sigma_\mu=0.2$, $\sigma_z=20$~\kms{}, $\sigma_\text{NL}=200$~\kms{}.
\end{enumerate}
I have chosen a fiducial $\Lambda$CDM cosmology with $\Omega_\text{M}=0.30$, $\Omega_\text{b}=0.04$, $\sigma_8(z=0)=0.84$, $H=80$~\kmsMpc, $n_\text{S}=1$. \virbius{} is run assuming the same cosmology, leaving free both the amplitude of the power spectrum, assuming that there are either {two or one specie(s)} of tracers, the zero-point calibration $\widetilde{H}$ and the parameters of the selection function. The grid used to compute the Fourier-Taylor algorithms {has $N=64$ elements per dimension}, which is sufficient to capture the details of $k=0.1$ Mpc$^{-1}$ and ensure a correct fast interpolation using the algorithm of Appendix~\ref{app:interpolation}.

\begin{figure}
	\begin{center}
		\includegraphics[width=.8\hsize]{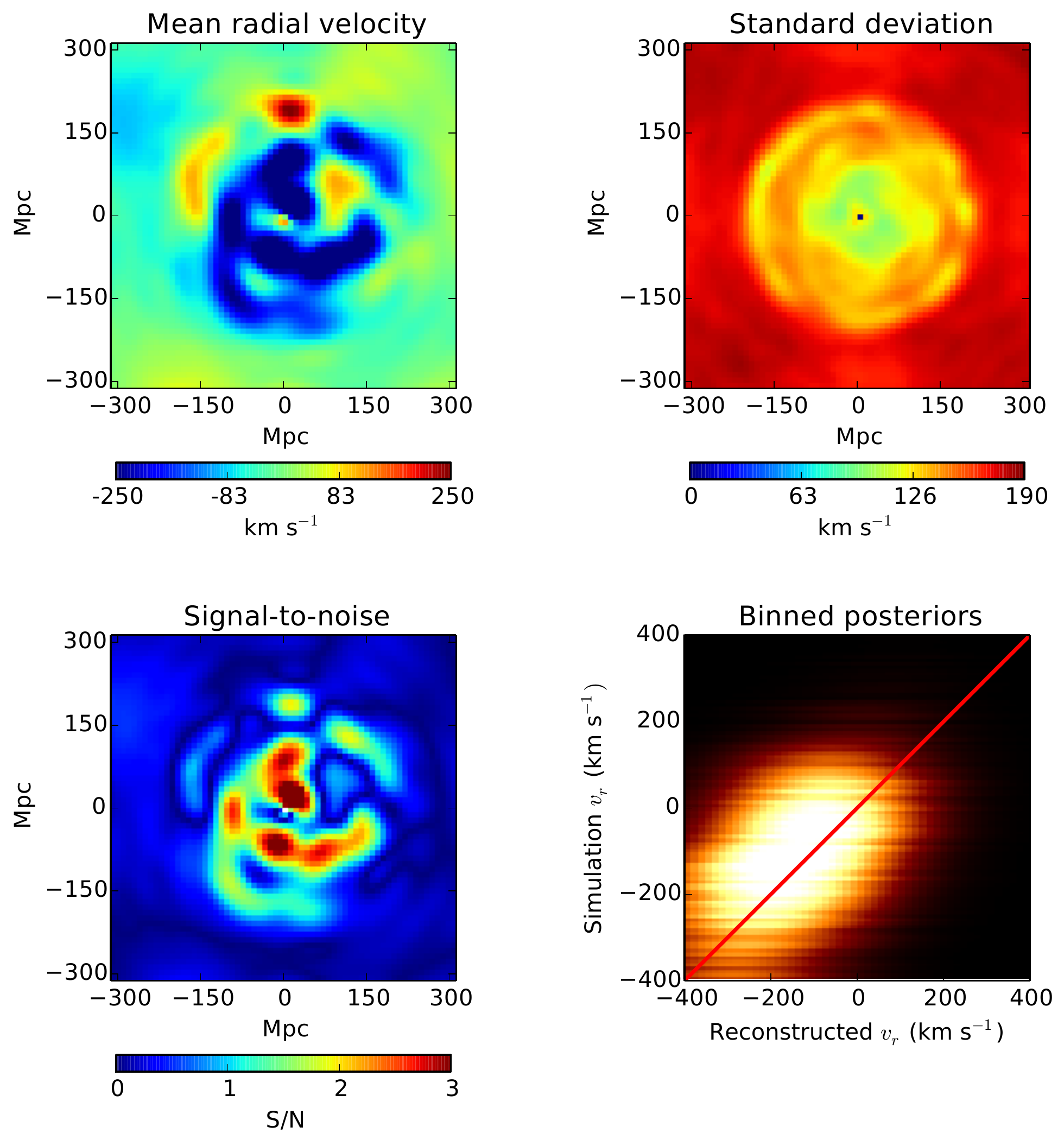}
   \end{center}
   \caption{\label{fig:grf_vr} Results of the test on the mock catalogue based on Gaussian random field:  {central slice of the} ensemble averages for the line of sight of the velocity field (top-left panel), the variance expressed as a standard deviation (top-right panel), the resulting signal to noise (bottom left panel).   {In the bottom right panel, I present} a Bayesian comparison between the mean reconstructed velocities and the actual true velocities of {the mock tracers} using the entire set of posterior distributions. }
\end{figure}

\begin{figure}
	\begin{center}
		\includegraphics[width=.8\hsize]{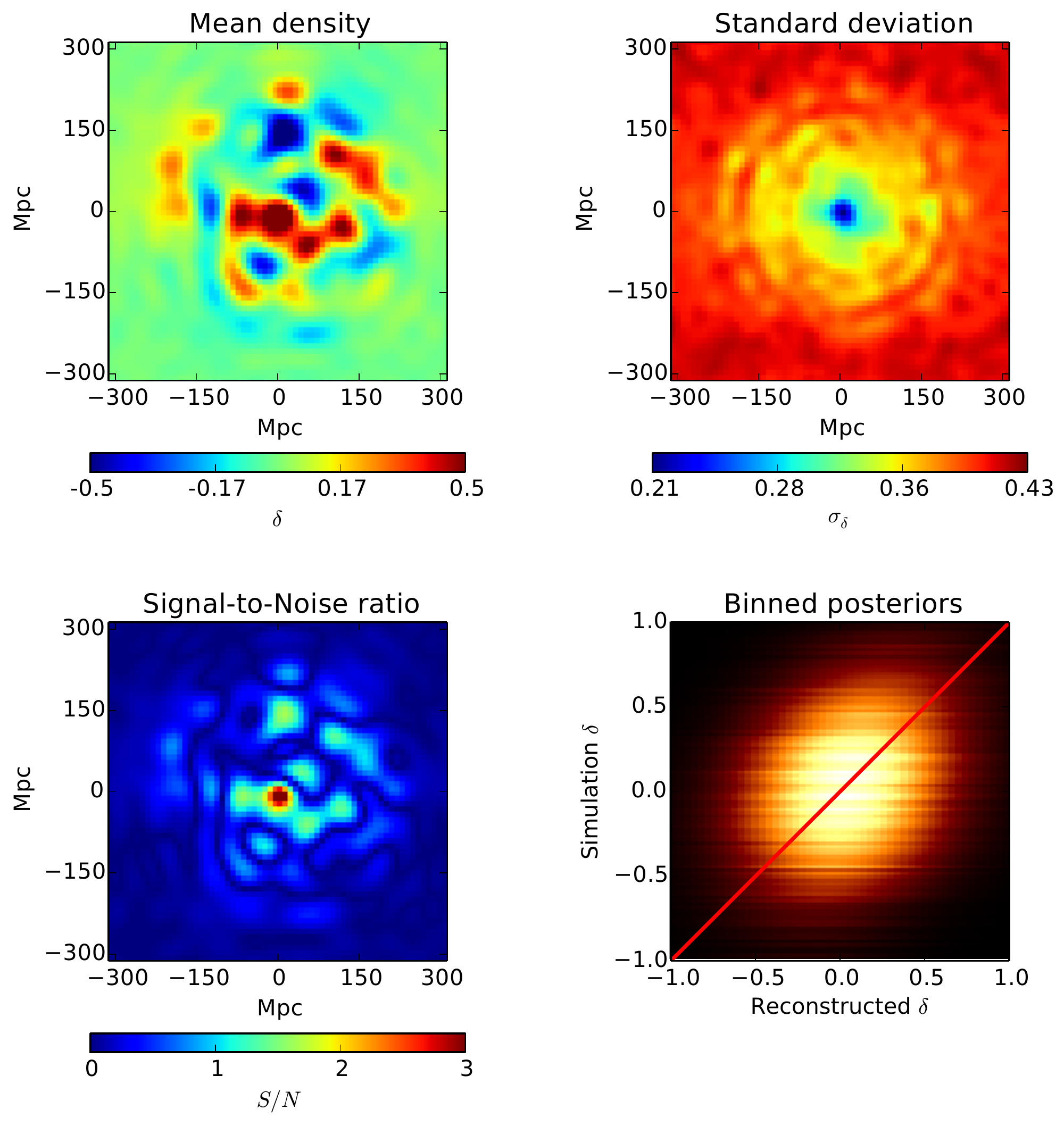}
   \end{center}
   \caption{\label{fig:grf_density} Results of the test on the mock catalogue based on Gaussian random field. All panels are related to density field distribution:  {central slice of the} ensemble average (top-left panel), the variance expressed as a standard deviation (top-right panel), the resulting signal to noise (bottom left panel). }	
\end{figure}

\begin{figure}
	\begin{center}
		\includegraphics[width=.8\hsize]{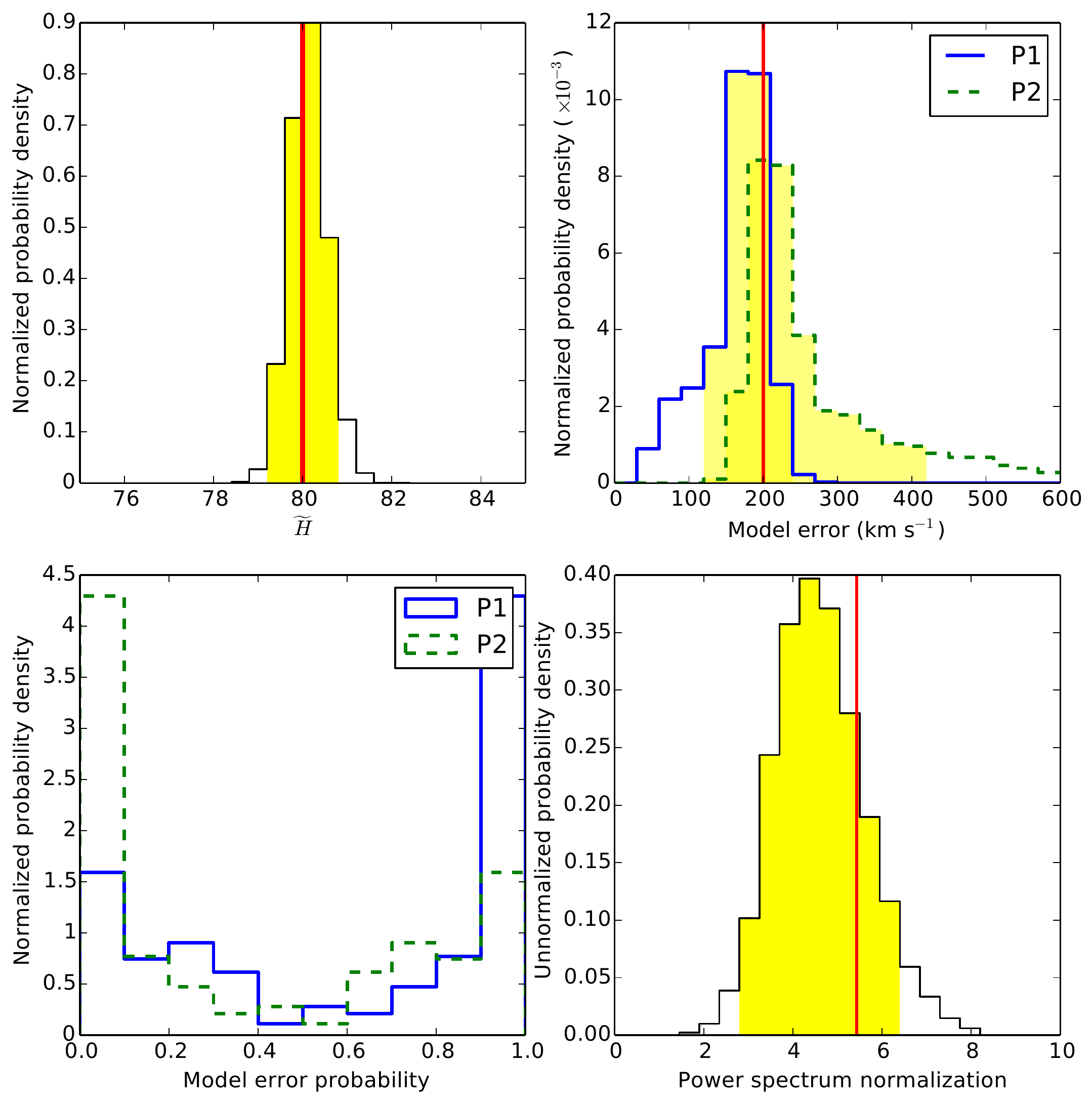}
   \end{center}
   \caption{\label{fig:grf_meta} Results of the test on the mock catalogue based on Gaussian random field. All panels show meta parameters posterior distributions:  the zero point calibration $\widetilde{H}$ (top left panel), the extra small scale non-linearities $\{ \sigma_{\text{NL},k} \}$ (top right panel), the probability of tracer type $\mathcal{P}$ (bottom left panel), and the power spectrum normalization in units of $10^6$~Mpc$^3$ (bottom right panel). The yellow shaded regions highlights the range of parameters for which the probability of containing the actual value is $\sim$90\%.}
\end{figure}

\begin{figure}
	\begin{center}
		\includegraphics[width=.8\hsize]{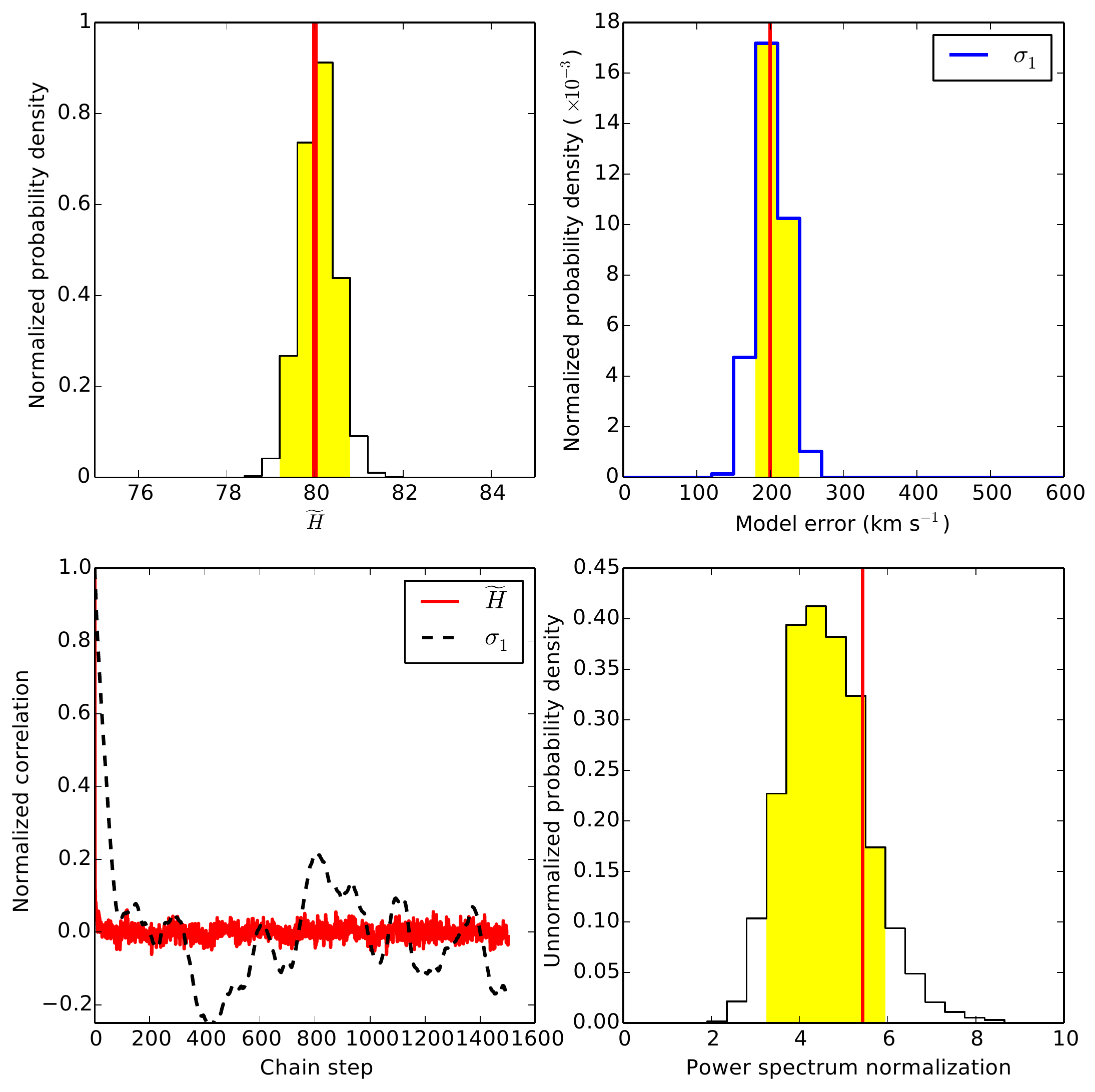}
   \end{center}
   \caption{\label{fig:grf_meta_1t} {Analysis of the same mock catalogue based on Gaussian random field as presented in Figure~\ref{fig:grf_meta} but this time restricted to a single type of tracer. The panels present similar quantities: the zero point calibration $\widetilde{H}$ (top left panel), the extra small scale non-linearities $\sigma_\text{NL}$ (top right panel), the power spectrum normalization $A_\text{S}$ in units of $10^6$~Mpc$^3$ (bottom left panel). Additionally the auto-correlation of the chains is given in the bottom left panels. The yellow shading also represents the range of parameters accepted at 90\% probability.} }
\end{figure}

\begin{figure}
	\begin{center}
		\includegraphics[width=.4\hsize]{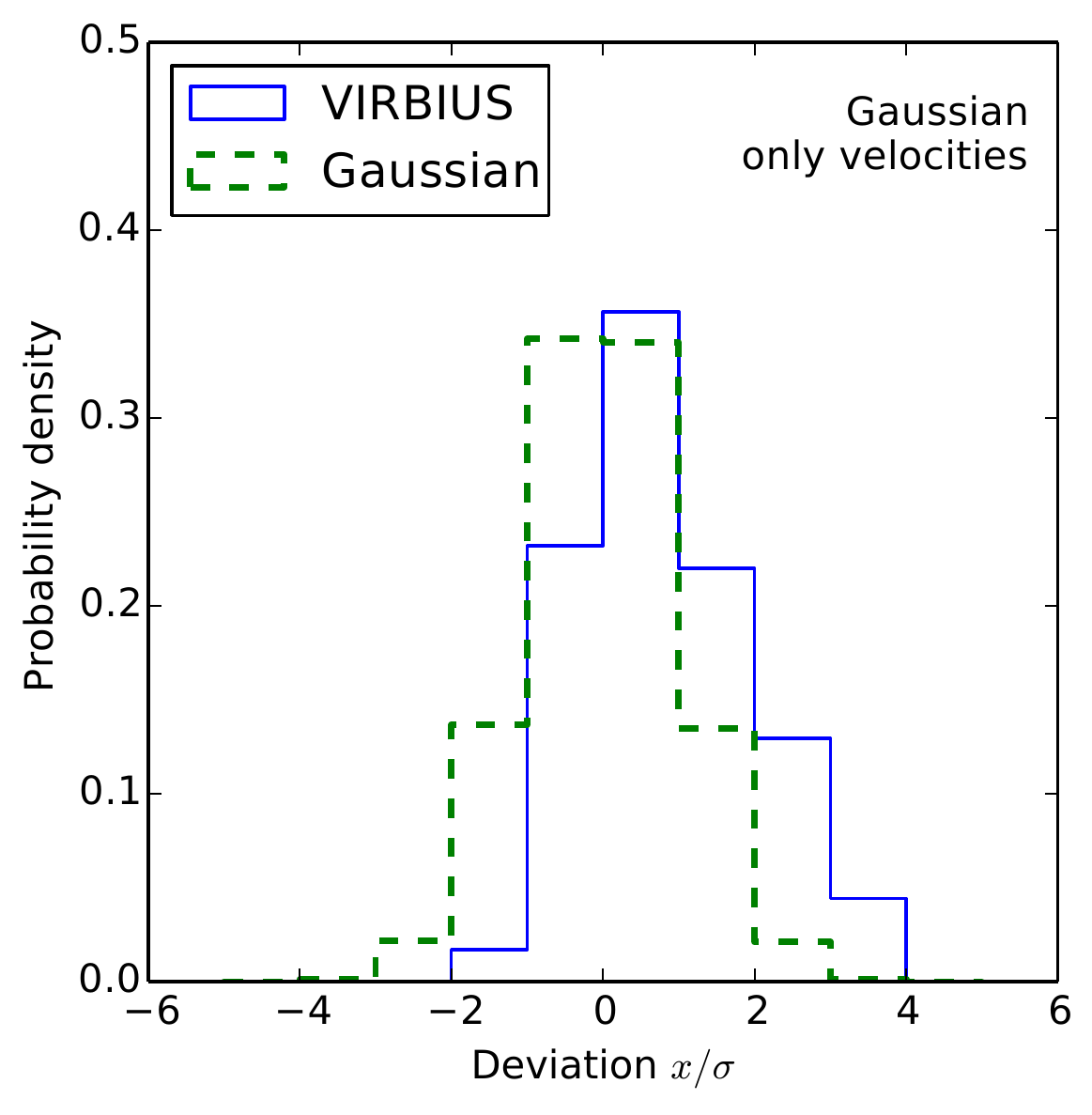}
		\includegraphics[width=.4\hsize]{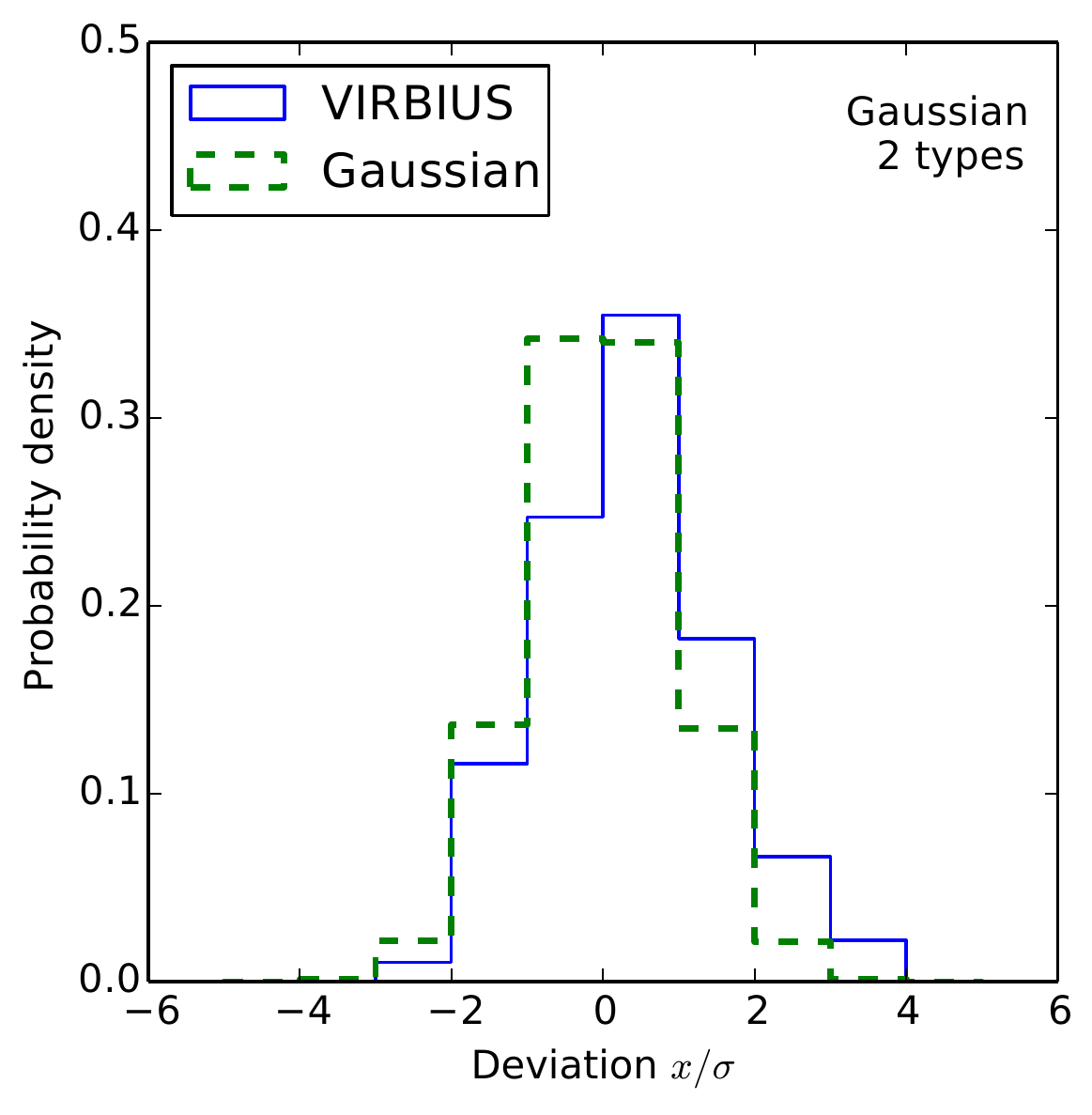}
	\end{center}
   \caption{\label{fig:residuals} Distribution of the velocity residuals $\epsilon_r$ between the simulation and the reconstructed mean component of the peculiar velocities for the different experiments on mock catalogues generated from Gaussian random fields. Left panel refer to the same experiment as Figure~\ref{fig:grf_precise_vr} and right panel to Figure~\ref{fig:grf_vr}.}
\end{figure}

Before considering the whole analysis, I am showing in Figure~\ref{fig:grf_precise_vr} and \ref{fig:grf_precise_density} the result of reconstructing the velocity field from an exactly known cosmology, distance and small scale non-linearities. I used a Gaussian noise with an amplitude of 200~\kms{} to model the small scale non-linearities as indicated above. In Figure~\ref{fig:grf_precise_vr} (Figure~\ref{fig:grf_precise_density} respectively), I am showing the reconstruction of the line-of-sight component of the velocity field (density field respectively). The figures were generated with {1055} Monte-Carlo samples.
In Figure~\ref{fig:grf_precise_vr}, the top left panel (top right respectively) shows the ensemble mean radial component (the standard deviation respectively) of the velocity field. The bottom left panel shows the signal-to-noise obtained by dividing the field of the top left panel by the one shown in the top right panel. Finally, the bottom right panel is obtained by showing all the individual posterior distribution for the velocities. These posterior distribution are binned on a grid and the total intensity is the sum of the posterior intensities for each considered grid element. The amplitude of the distribution is coded in color with respect to the $x$-axis, while the line-of-sight component of the velocity of the corresponding object as given by the simulation is shown on the $y$-axis. If the algorithm works correctly, all the distributions should overlap with the thick red diagonal, {though not necessarily centered as the distributions are correlated. The correlation actually removes the effective number of independent samples compared to the number of elements that are plotted. The unbiased aspect of the method is exhibited by the plots of the normalized residuals in Figure~\ref{fig:residuals}, where I have represented the histograms of the quantity
\begin{equation}
	\epsilon_r = \frac{\bar{v}_r(\mvec{x})-v^\text{sim}_r(\mvec{x}))}{\sigma(\mvec{x})}.
\end{equation}
The same histogram for a pure Gaussian distribution is represented  with dashed green line. It can be immediately seen that the overall distribution of residuals is close to Gaussian with unit variance. Additionally, there is no obviously strong shift in the mean. }

The panels are similar in Figure~\ref{fig:grf_precise_density} but this time for the density field.
We see that in this optimistic configuration the velocity field is very well reconstructed, on the other hand the density field is already significantly noisy. These results are the benchmarks to which we will compare the performance of other reconstructions.

The results for more complete tests are given in Figure~\ref{fig:autocorr_chain}, \ref{fig:grf_vr}, \ref{fig:grf_density}, \ref{fig:grf_meta} and \ref{fig:grf_meta_1t}. The figures were generated with {6894 (3037 respectively)} Monte-Carlo samples {for the two types (one type respectively) scenario}.  Figure~\ref{fig:autocorr_chain} shows the convergence rate of some meta parameters along the Markov chain (top left panel for this mock catalog). The convergence for $\widetilde{H}$ is typically fast, decorrelating in a few iterations. That is the result of the partially collapsed Gibbs sampler. The other shown parameters, related to the Gaussian mixture of Section~\ref{sec:sigmaNL_fit}, take substantially more steps to decorrelate, {of the order of a few hundreds}. This is expected as they are tied to a particular realization of the velocity field, contrary to other meta parameters which are obtained through direct marginalization over the velocity field. We can expect the convergence length to increase when the number of tracer is low or the noise is high as the uncertainty over the velocity field will increase and thus it takes more steps to explore the parameter space. We will see how it is the case with the other mock catalogues in the next section.

Figure~\ref{fig:grf_vr} gives a synthetic view of the velocity field component of the posterior distribution. The details of the panels have been given previously for Figure~\ref{fig:grf_precise_vr}. Comparing Figure~\ref{fig:grf_precise_vr} to \ref{fig:grf_vr} shows how much the relaxation of the assumption of fixed cosmology, noise and distances degrades the quality of the reconstructed velocities.

Figure~\ref{fig:grf_density} gives a similar view as for the velocity field, but this time for the density, i.e. the divergence of the velocity field. The view is the same as the one in Figure~\ref{fig:grf_precise_density} but this time we have all the parameters being sampled. This field is expected to be much noisier as it is indeed the case when looking at ensemble average quantities and signal-to-noise (S/N). In the centre, where S/N is the highest we are only at maximum at 3-$\sigma$, while for peculiar velocities it was possible to go at higher S/N. The resulting comparison of the individual posterior distribution in the bottom-right panel shows this high uncertainty. Compared to Figure~\ref{fig:grf_precise_density} there has been a strong loss of information on the density field. It reaches a point where the binned posterior in the bottom right panel shows no constraint at all from data.

In Figure~\ref{fig:grf_meta}, I show the individual distributions of the meta parameters of the chain, i.e. the effective Hubble constant $\widetilde{H}$, the model error amplitude, the probability for each type of model error, and the overall power spectrum normalization $A_S$. For $\widetilde{H}$, $A_S$ and the model error, a thick vertical red line shows the value used to generate the mock catalogue. {I am also showing in shaded yellow the range of values compatible at 90\% with the data according to the model.} As expected all distributions {are compatible} with the thick vertical line at their peak positions, i.e. always well within the {90\% region} of the distribution. In the case of the model error amplitude (upper right panel), we are faced with two distributions overlapping with the fiducial value.  Finally the posterior distribution shown in the bottom left panel do not indicate a strong imbalance between the different error model, which exactly corresponds to the significant overlap of the distributions in the top right panel.  {The results of the distributions obtained assuming a single type are given in Figure~\ref{fig:grf_meta_1t}. All the meta distributions are of course narrower than for the test with two types, especially for the distribution of $\sigma_\text{NL}$. The width of the distribution for $A_S$ is also visibly narrower. The peculiar velocity field, not represented here, is however mostly unaffected. In the test with two assumed types of tracers, the algorithm separated the data in two pieces: one including more than 93\% of the tracers 50\% of the time, and the others. This separation results in a small number of objects having high velocity dispersion, which both skew the P1 distribution towards low value of $\sigma_\text{NL}$ and allow a the P2 distribution to venture to very high values of this same $\sigma_\text{NL}$. No obvious bias is visible in all the distributions. The autocorrelation of the chain is shorter by a factor $\sim$2. } {These tests indicate that} the algorithm and the software works as expected on an idealistic test case.

\begin{figure}
	\begin{center}
		\includegraphics[width=.8\hsize]{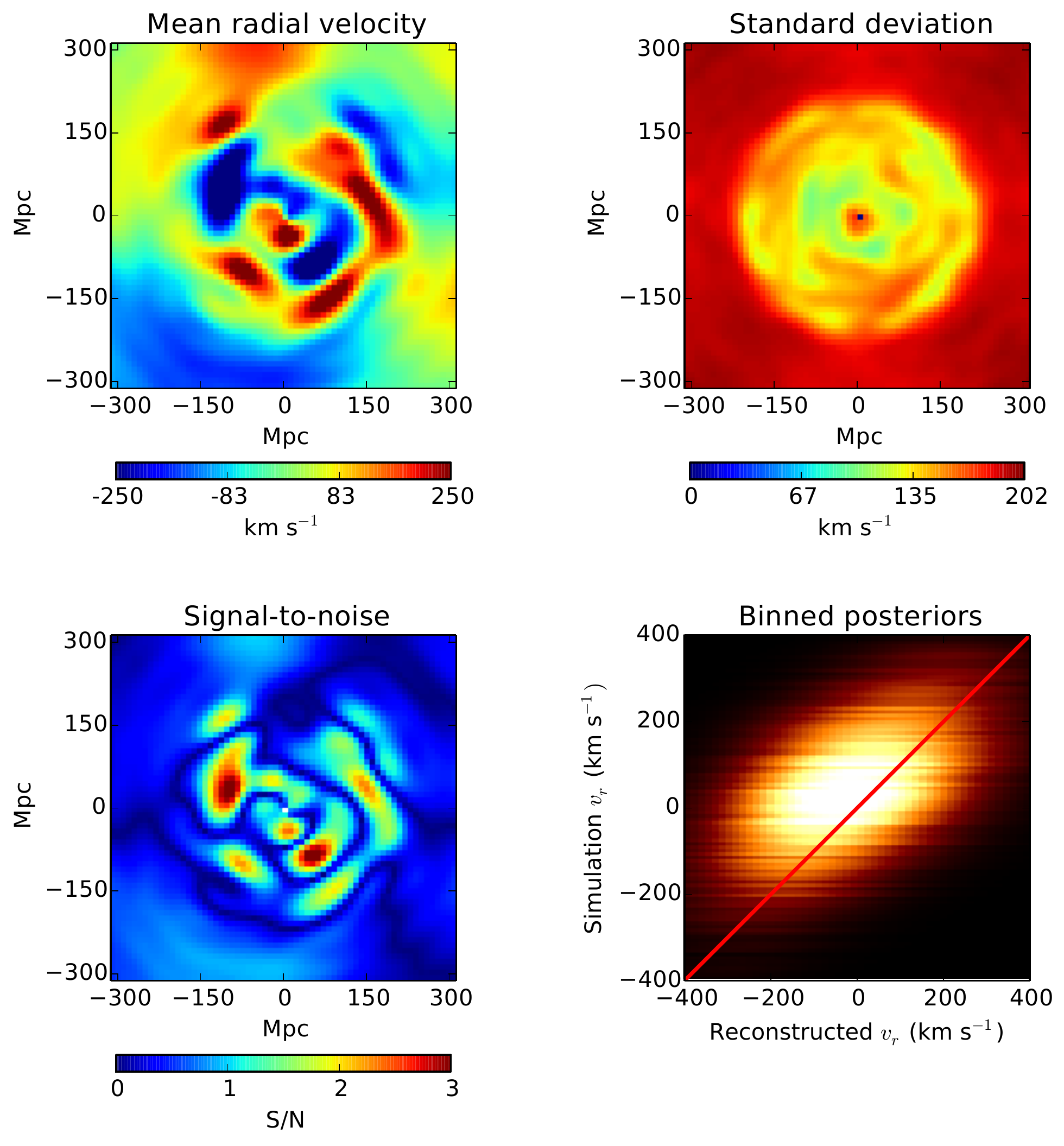}
   \end{center}
   \caption{\label{fig:h3000_vr} Same as Figure~\ref{fig:grf_vr} but for the H3000 mock catalogue.  }
\end{figure}

\begin{figure}
	\begin{center}
		\includegraphics[width=.8\hsize]{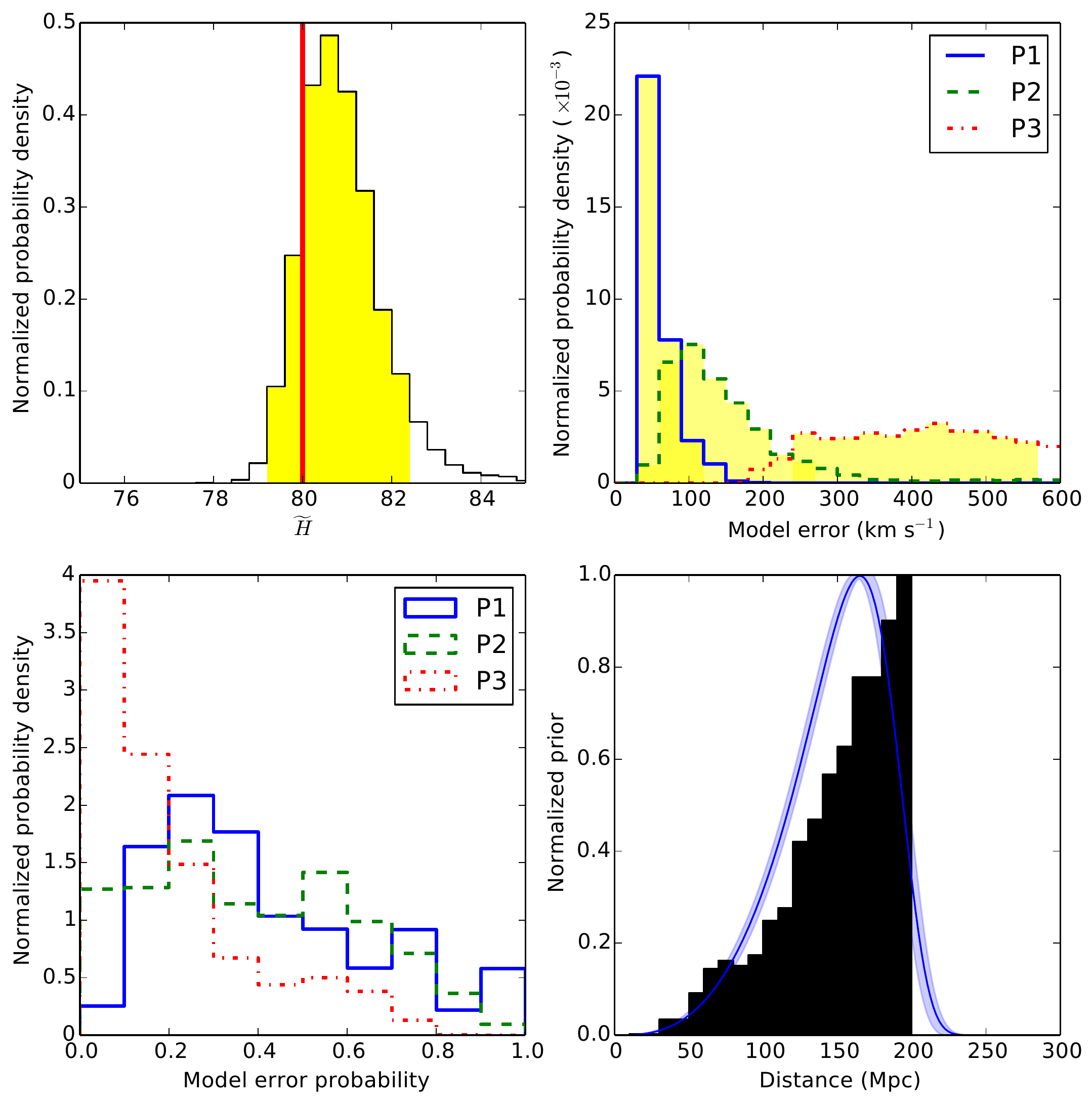}
   \end{center}
   \caption{\label{fig:h3000_meta}  Results of the test on the the H3000 mock catalogue:  meta parameters distributions:  $\widetilde{H}$ (top left panel), $\{ \sigma_{\text{NL},k} \}$ (top right panel), $\mathcal{P}$ (bottom left panel), selection function (bottom right panel). Some bins are still clearly dominated by the noise due to the long correlation length of the chain.}
\end{figure}

\begin{figure}
	\begin{center}
		\includegraphics[width=.8\hsize]{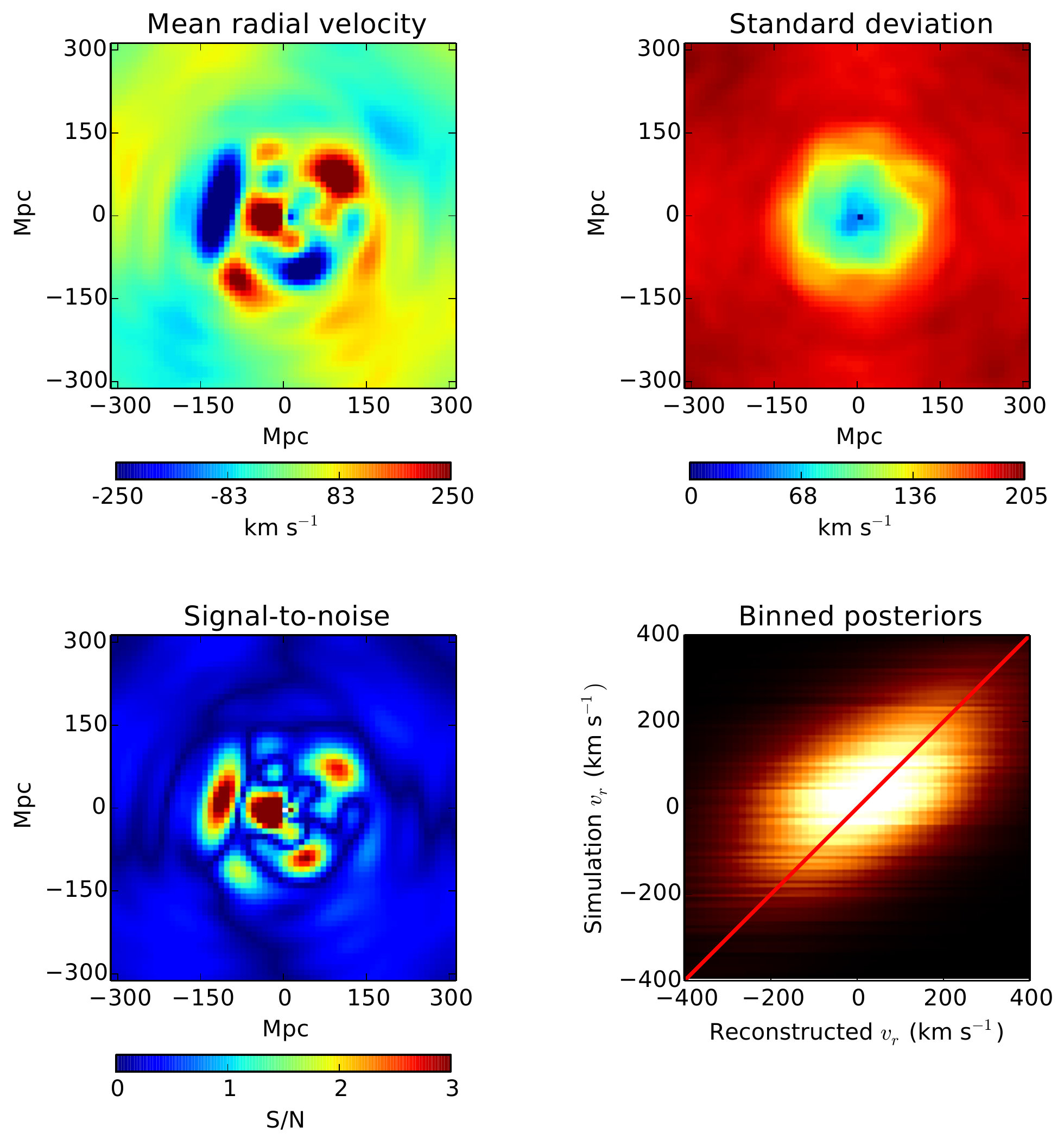}
   \end{center}
   \caption{\label{fig:hcomplex_noisy_vr} Same as Figure~\ref{fig:grf_vr} but for the Hcomplex mock catalogue.}
\end{figure}

\begin{figure}
	\begin{center}
		\includegraphics[width=.8\hsize]{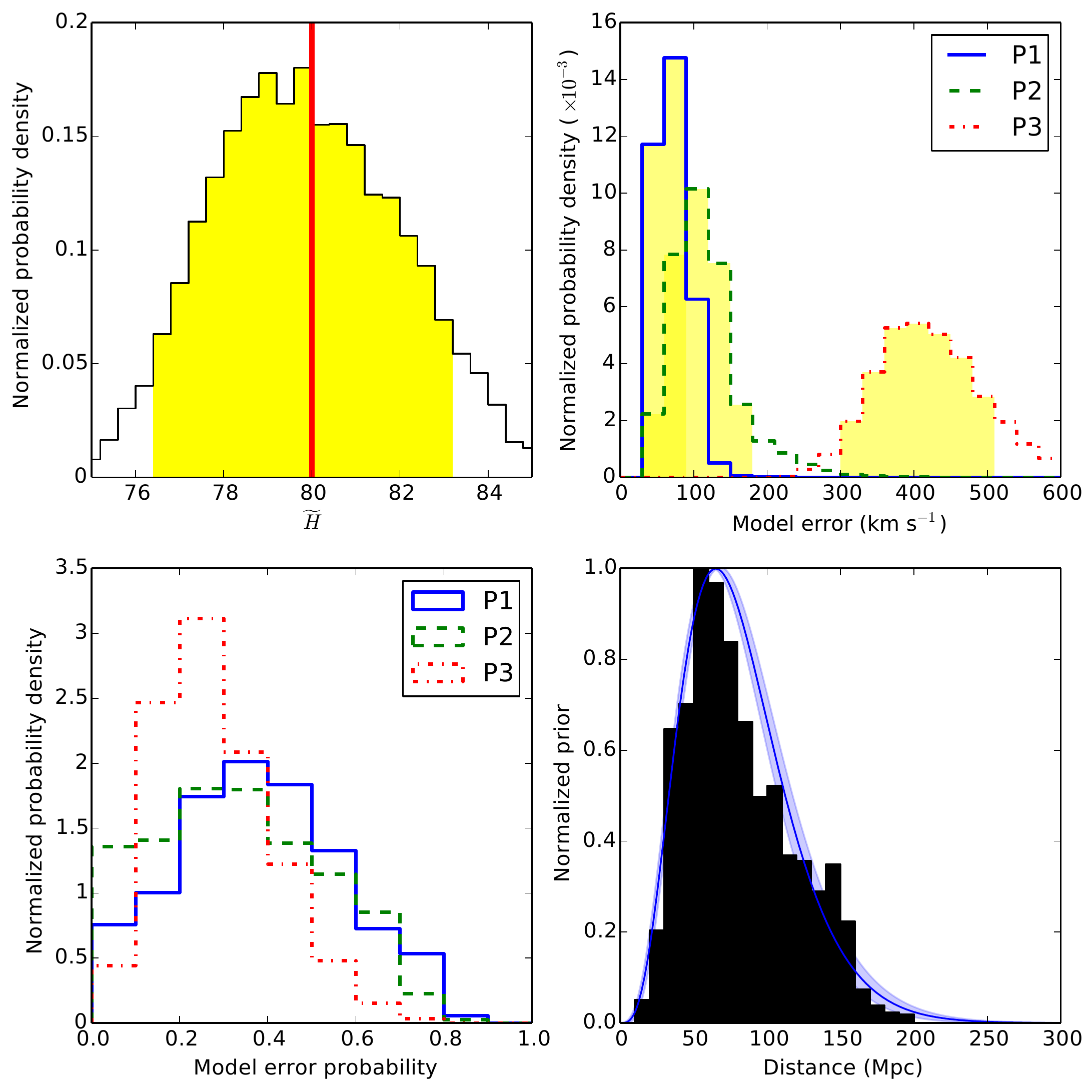}
   \end{center}
   \caption{\label{fig:hcomplex_noisy_meta} Same as Figure~\ref{fig:h3000_meta} for the Hcomplex mock catalogue.}
\end{figure}

\subsection{Halo based mock catalogue}
\label{sec:test_mock_halo}

In this Section, I consider more realistic, but more complicated mock catalogue to which I apply the methodology developed in this article. Two mock catalogues are considered, both based on the halos of a cosmological pure dark matter $N$-body simulation. This simulation have been computed using the following cosmological parameters: $\Omega_\text{M}=0.30$, $\Omega_\text{b}=0.045$, $\Omega_\Lambda=0.70$, $H = 80$~\kmsMpc, $\sigma_8 = 0.80$. The volume covered by the simulation is a cube with a side of 500\Mpch{} with $512^3$ particles.  From this simulation, halos were extracted using the {\tt\sc Rockstar} software \citep{rockstar}. The minimum halo size have been kept to its default value of 10, checking that particles are effectively bound. The total number of halos of the simulation is thus 238~520. The two catalogues are created from the same simulation but choosing different selection properties.  The first mock catalogue, nicknamed H3000, is generated by extracting randomly 3000 halos from the rockstar catalog at a distance less than 200\Mpch{} of the center of the simulation box. The chosen distance modulus error is $\sigma_\mu=0.2$, typical for a Tully-Fisher relation. The second mock catalogue, nicknamed Hcomplex, is built from a more complex selection function, derived from the $M/L$ relation used in \cite{Lavaux08}, equation (7). The principal condition for acceptance in the catalogue is to have $L / (d^L)^3 \le 2.78\times 10^4$, with $L$ in solar luminosities and $d$ the comoving distance in Mpc/h. The selection is not supposed to be strictly realistic, but to mimic a realistic abundance for distance catalogue. This mock catalogue has 2000 tracers, with a distance modulus error $\sigma_\mu = 0.1$, typical of higher quality distance indicator like Supernovae (SNe) or Tip of the Red Giant branch (TRGB). 

For these mock catalogues I restrained from fitting the amplitude of the power spectrum $A_S$ at the same time as the other parameters. The problem comes from the degeneracy between $A_S$ and the distribution of tracers of the velocity field. In halo catalogues these two quantities are not independent as tracers are typically located in the peak of the density distribution. This tends to credit the velocity field with more power to try to homogenize the distribution of tracers. This problem arises when we are faced with data which needs a sufficiently precise prior on the density of tracers. {I have run a chain for which the selection is exactly known, i.e. homogeneous in luminosity distance space, but for which $A_\text{S}$ is left free. The mean recovered value is $2.9\pm 0.5$ times higher that the value used to make initial conditions. Thus} I postpone the resolution of this problem to future work. Here, I will limit myself to fix $A_S$ to its fiducial value and investigate the rest of the parameter space. 

The results are given in Figure~\ref{fig:autocorr_chain}, \ref{fig:h3000_vr} and \ref{fig:h3000_meta} for the H3000 halo mock catalogue, and in Figure~\ref{fig:hcomplex_noisy_vr} and \ref{fig:hcomplex_noisy_meta} for the Hcomplex halo mock catalogue. The length of the chain is 10824 for H3000, and 3905 for Hcomplex. The convergence test is shown in Figure~\ref{fig:autocorr_chain} (top right for H3000, bottom middle for Hcomplex). The convergence of the parameters of the Gaussian mixture is quite slow for H3000 due to the substantial uncertainty on the velocity field potential. This triggers a large correlated exploration of distances and model error parameters. The chain concerning H3000 has 10825 samples, whereas the one concerning Hcomplex has 1796 and is clearly converged. However for Hcomplex, this convergence is much faster, confirming the previous scenario. In that case the tracer density is sufficiently high to largely reduce velocity field uncertainties in the effective volume covered by the catalogue. {That statement can actually be generalized. If the catalog of tracer is dense, then the velocity potential will be quite tightly constrained which does not leave much freedom for the auxiliary meta parameter like $\Sigma_\text{NL}$, which in turn reduce possibilities for the reconstructed true distances.}

The panels of the other figures are showing the same quantities as the ones in Figures~\ref{fig:grf_vr} and \ref{fig:grf_meta}, with the exception of the bottom right panel of Figures~\ref{fig:h3000_meta} and \ref{fig:hcomplex_noisy_meta} which shows the selection function of galaxies in the catalogue. In these same panels, I show the actual distribution of the galaxies for H3000 (a simple $d^2$ law) and Hcomplex as a function of luminosity distance. I note that the selection function for H3000 has been fitted by VIRBIuS according to prescription. However as there is a sharp cut at 200~Mpc in the mock catalog it is not able to reproduce this fairly. The parameter most affected by this discrepancy is in principle $\widetilde{H}$. We see that, within error bars, it does introduce significant bias. On the other hand, the galaxy population of Hcomplex is fairly represented by the selection function fitted by VIRBIuS (Figure~\ref{fig:hcomplex_noisy_meta}), and $\widetilde{H}$ is measured without any systematic effect.

The application of this Bayesian methodology on the two mock catalogues is satisfactory. All the posterior distributions are in agreement compared to the velocities of the simulations and the meta parameters are in agreement with the input value, such as $\widetilde{H}$ and the selection function. The analysis also yields that the halos can be classified into two main populations: the results on the measurement of $\Sigma_\text{NL}$, assuming three populations, is quite clear in particularly for the top right panel of Figure~\ref{fig:hcomplex_noisy_meta}. There is a population of halos with a model error of $\sim 400$~\kms and another at $\sim 100$~\kms. The probability assigned to each model is not clear, all models getting a share of $\sim$30\%. Which means a probability of $\sim$66\% for the low velocity dispersion model and $\sim$33\% for the high velocity dispersion one.  {It is possible to run a meta-analysis for which the chain is run assuming different maximum number of populations. The likelihood of the data, marginalized according to sampled parameters, given the population model can then be computed alongside the Bayes factor between the different models. }

\begin{figure}
	\begin{center}
		\begin{tabular}{cc}
			H3000 & Hcomplex \\
			\includegraphics[width=.4\hsize]{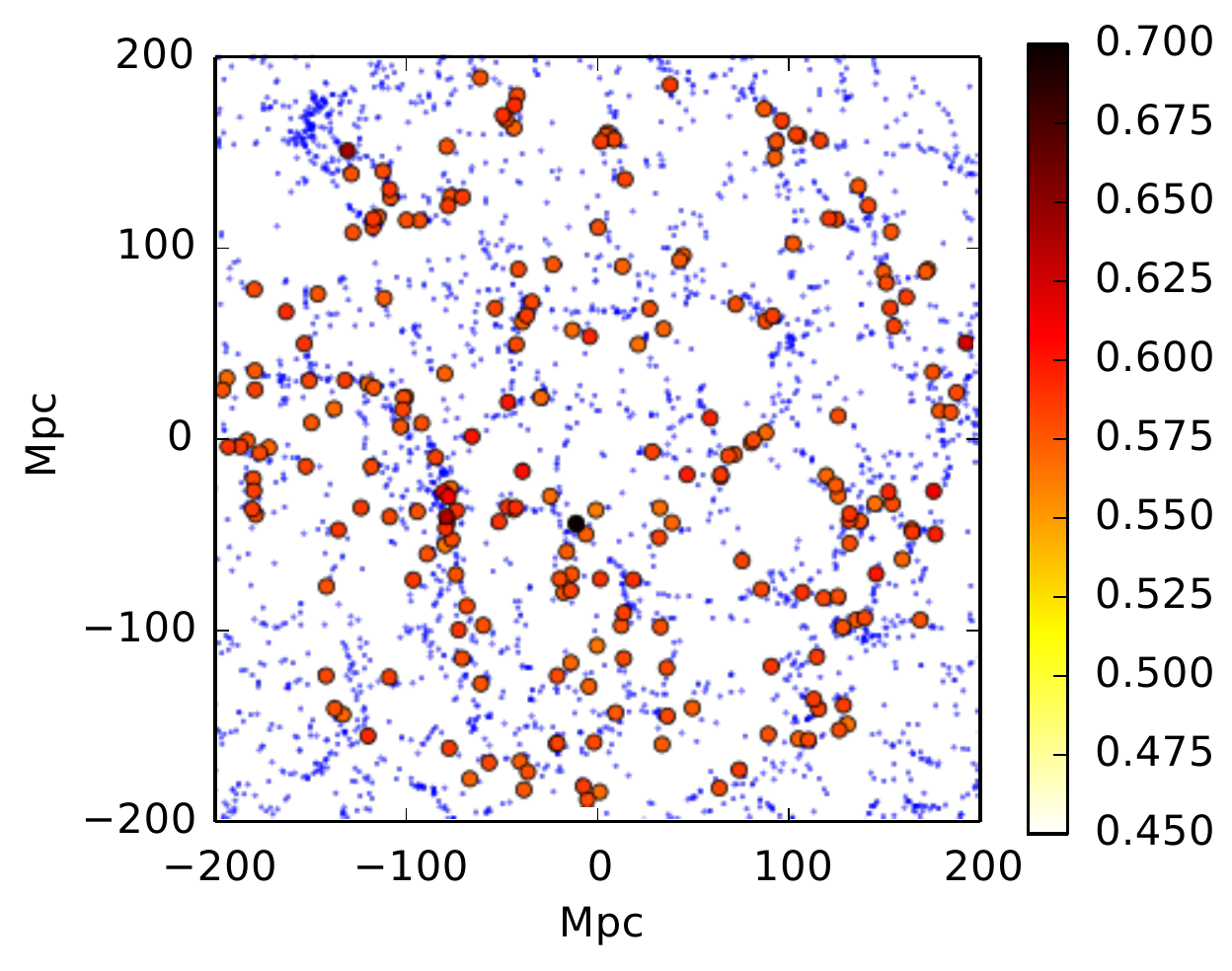} &
			\includegraphics[width=.4\hsize]{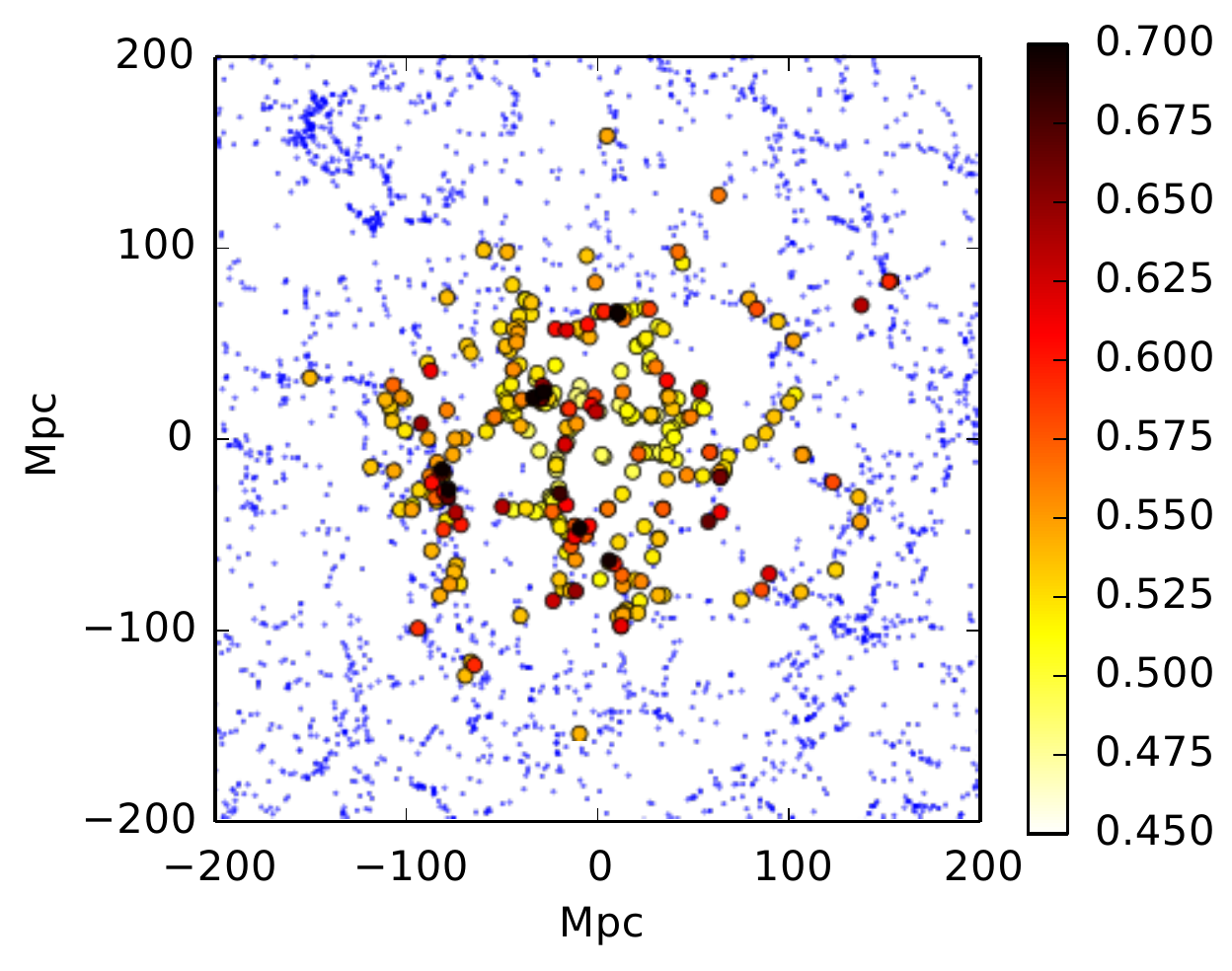}
		\end{tabular}
   \end{center}
   \caption{\label{fig:classification} Results of the automatic classification for the two halo mock catalogues (H3000 in the left panel, Hcomplex in the right panel). I am showing a thin slice of both the mock catalogues and the simulations. The blue points correspond to the halos of the original simulation while the discs correspond to the mock catalogues. The disc are coloured according to the probability that a tracer has a velocity dispersion greater than 100~\kms. }
\end{figure}

This shows the power of this Bayesian methodology: in addition to allowing a consistent reconstruction of peculiar velocities, it is now possible to quantitatively measure the probability that a halo (or a galaxy in real distance catalogues) belongs to a high velocity dispersion population or not in observations. This can be the case if the halo hosting the galaxy is a substructure for example. It could also be a supernova deeply embedded in the galaxy potential well. In Figure~\ref{fig:classification}, I show an attempt at classifying the halos depending on their probability of belonging to one or the other population. In this Figure a slice of the two catalogues are shown: in the left panel H3000, and in the right panel Hcomplex. The halos of the simulation are represented as small blue points. The halos selected to be part of the mock catalogues are shown as coloured discs. The colour of the discs is assigned depending on whether the probability that it has a high velocity dispersion, i.e. greater than 100~\kms. 
While in the H3000 catalogue it is difficult to separate two populations, in the Hcomplex catalogue two populations clearly emerge spatially. The darkly coloured discs are qualitatively more likely to be at knots of the cosmic web, compared to the lightly coloured discs.
{Of course, as indicated early on in Section~\ref{sec:stats}, the other main interest of the classifier consists in the natural removal of potential outliers from the observational datasets. Typically the outliers will be put in separate type with a huge velocity dispersions which will remove any effect that they could have on the fitting of the velocity field. }

\section{Conclusion}
\label{sec:conclusion}

The advent of new large distance catalogues, like 6dFv \citep{Campbell14} or Cosmic Flows-2 \citep{Tully13} is opening a new era in cosmic velocity field analysis. This era will continue with TAIPAN \citep{Beutler11} and WALLABY \citep{Duffy12}. To face this avalanche of new data and get the most of them, new methods of analysing redshift and distance surveys are required. 

For this work, I develop, implement and test a Bayesian algorithm to reconstruct peculiar velocity field from distance catalogues. This method is based on a Bayesian formulation of the reconstruction problem, assuming that the data follows the model given in Equation~\eqref{eq:model}. Additionally, all cosmological expansion corrections are already built in. I have shown that in the case of mock catalogues following perfectly the data model, then the algorithm manages to recover all quantities in an unbiased way. However, for more realistic mock catalogues taken from $N$-body simulations, some bias is introduced due to approximation either in the model, in the used priors for distances and velocities. This bias is the most prominent for the amplitude of the power spectrum when the noise on distances is sufficiently large to have to rely on velocity field correlations to infer the position of tracers. {Finally, this algorithm includes an automatic tracer classifier that makes it interesting to classify object: either as observational outliers, or as object belonging to collapse structures and thus exhibiting higher velocity dispersions.}

While the results obtained here are promising, they are not the end of the development of this method. One of the next step would be to harden the fitting of the amplitude $A_S$ of the density fluctuation power spectrum by adding a bias field in the distance prior. This bias field would need to be stochastically related to $\Delta \Theta \sim -\delta_\text{m}$ as it is more likely to have halos in locations where the flow {is} converging. However the sampling of this bias field would be substantially more complex and I am postponing this to later work. Simpler improvements would consist in allowing several samples with different radial selection function and fitting the inverse Tully-Fisher relation at the same time as the other parameters. Finally the density field sampled using this algorithm could be used as a prior for either ARES \citep{jasche_methods_2013} or BORG \citep{JASCHEBORG2012,JASCHE2014Borgsdss} algorithms of reconstruction of density field from galaxy spectroscopic surveys.

\section*{Acknowledgments}

I would like to thank Michael J. Hudson for many useful discussions and encouragements to pursue this project when it was in its early stage. I also would like to thank Jens Jasche for many useful discussions on Bayesian statistics. 
I thank Stephen Turnbull for reading an early version of this manuscript.  {I thank the anonymous referee for the detailed reading and the suggested corrections which have greatly improved the manuscript.}

I acknowledge support from CITA National Fellowship and financial support from the Government of Canada Post-Doctoral Research Fellowship. Research at Perimeter Institute is supported by the Government of Canada through Industry Canada and by the Province of Ontario through the Ministry of Research and Innovation.

This work was granted access to the HPC resources of The Institute for scientific Computing and Simulation financed by Region \^Ile-de-France and the project Equip@Meso (reference ANR-10-EQPX-29-01) overseen by the French National Research Agency (ANR) as part of the ``Investissements d'Avenir'' program.

This work has been done within the Labex ILP (reference ANR-10-LABX-63) part of the Idex SUPER, and received financial state aid managed by the Agence Nationale de la Recherche, as part of the programme Investissements d'avenir under the reference ANR-11-IDEX-0004-02.

Special thanks go to St\'ephane Rouberol for his support during the course of this work, in particular for guaranteeing flawless use of all required computational resources.

I acknowledge financial support from "Programme National de Cosmologie and Galaxies" (PNCG) of CNRS/INSU, France.


\clearpage
\onecolumn
\appendix
\section{Adaptive sampling from a non-trivial one-dimensional distribution}
\label{app:1d_sample}

The canonical algorithm to sample from an arbitrary probability distribution is based on the cumulative probability function:
\begin{equation}
	P(x \leq X) = \int_{-\infty}^{X} \text{d}x\; p(x).
\end{equation}
If $y$ is sampled from a uniform distribution bounded by $[0,1[$, then
\begin{equation}
	x = P^{-1}(y),
\end{equation}
where $P^{-1}$ is the inverse function of $P$, is distributed according to $p$.
Unfortunately in the cases of this work, none of the distributions are analytical, and even less integrable and invertible. We must rely on a numerical scheme for this. I consider different limiting cases to try to optimize the random number generation. 

\subsection{Single modal sharply peaked or weakly multi-modal}

The first case that I consider is in the approximation that $p(x)$ has mostly a single sharp peak, with maybe some smaller insignificant sub peaks around this one. This is typically the case when sampling the cosmological hyper parameters like the Hubble constant $H$, the distance calibration $\widetilde{H}$ or the spurious noise on velocities $\sigma_\mathrm{NL}$. The evaluation of the posterior in these cases is costly, involving the inversions of large matrices with a size of the order the number of tracers in the original catalogue. It is thus required to make as few evaluations as possible. The position and sharpness of the peak is not clearly a priori known and we have to rely on adaptivity. The algorithm proceeds as follow:
\begin{enumerate}
\item It looks for the maximum $x_\text{max}$ of the 1D posterior distribution using the Brent minimization algorithm specified in the GSL. The evaluate of the posterior are cached as the minimization proceeds.
\item From the cached values, it estimates the curvature by computing the average of $(x - x_\text{max})^2$ from the cached values, using a linear interpolation of the logarithm of the posterior.
\item This curvature gives the typical sampling size required to capture the details of the posterior, which is now sampled regularly relative to $x_\text{max}$. The evaluation is executed till the ratio between the edge value of the posterior and its peak value is less than some threshold. This procedure is run in parallel on shared memory computers.
\item The final sample is then generated using a rejection sampler, assuming the evaluated discrete distribution as proposal distribution.
\end{enumerate}

This algorithm is robust and manages to generate sample in cases the distribution extends over several orders of magnitudes, while retaining accuracy of the target distribution. This part is specifically critical to ensure the stability of the final Markov Chain.

\subsection{Multi modal smoothly varying}

The second case corresponds to the sampling of distances assuming a fixed large scale velocity field. Many distance posterior will be simple. However some will be truncated (for example if we are on the edge of the survey), or severely multi-modal in the case of a collapsing structure. For that reason we cannot entirely apply the previous algorithm and we in place simplify it by removing the steps 1 and 2. In place I just evaluate regularly the distance posterior in a sufficiently fine-grained way to resolve the evolution of the velocity field and the typical error bar on distance inferred from both $\sigma_\text{NL}$ and the distance modulus error. The rejection sampler is then used again to clean the discreteness effects. This algorithm can only be used here because the evaluation of the posterior is very fast, contrary to the most cases considered in the previous section.

\section{Fourier-Taylor interpolation on a non regular mesh}
\label{app:interpolation}

The statistical algorithm described in this work makes use of tracers put at positions not located on a mesh. It is notably necessary to compute $\mvec{\Psi}(\mvec{r})$ at any location in the volume of reconstruction. Of course $\mvec{\Psi}(\mvec{r})$ is obtained through a Fourier synthesis operation described in Equation~\eqref{eq:psi_fourier} that we take now as a definition, in dimension $d$:
\begin{equation} 
 \mvec{\Psi}(\mvec{r}) = \frac{1}{L^d} \sum_{q=1}^{N_q} \frac{i \mvec{k}_q}{|\mvec{k}_q|^2} \text{e}^{i \mvec{k}_q . \mvec{r}} \hat{\Theta}(\mvec{k}_q).
\end{equation}
We make use of the idea originally developed by \cite{AndersonDahleh96} to have a fast interpolation of Fourier series. The Taylor expansion of trigonometric functions converges very rapidly, it is thus tempting to do a Taylor expansion of $\mvec{\Psi}(\mvec{r})$ against the nearest grid point $\mvec{g}$. The general expansion is the following:
\begin{align}
 \Psi_s(\mvec{r}) &= \frac{1}{L^d} \sum_{q=1}^{N_q} \frac{i k_s}{|\mvec{k}_q|^2} \text{e}^{i \mvec{k}_q . (\mvec{r}-\mvec{g})} \hat{\Theta}(\mvec{k}_q) \text{e}^{i \mvec{k}_q . \mvec{g}} \\
  & = \frac{1}{L^d} \sum_{q=1}^{N_q} \sum_{n=0}^{+\infty}  
   \frac{i^n}{n!} \left(\sum_{j=1}^d (\mvec{k}_q)_j (r_j-g_j)\right)^n
 \frac{i k_s}{|\mvec{k}_q|^2} \hat{\Theta}(\mvec{k}_q) \text{e}^{i \mvec{k}_q . \mvec{g}} \\
   &= \frac{1}{L^d} \sum_{q=1}^{N_q} \sum_{n=0}^{+\infty}  
   \frac{i^n}{n!} \left( \sum_{\{a_p\} } \prod_{p=1}^d \left(r_j-g_j\right)^{a_p} F(\mvec{k}_q,n,\{a_p\})\right)
 \frac{i k_s}{|\mvec{k}_q|^2} \hat{\Theta}(\mvec{k}_q) \text{e}^{i \mvec{k}_q . \mvec{g}} \\
 &= \sum_{n=0}^{+\infty} \frac{1}{n!} \sum_{\{a_p\} } \left(\prod_{p=1}^d \left(r_p-g_p\right)^{a_p}\right) Q_{s,n}(\{a_p\},\mvec{g}),
\end{align}
with 
\begin{align}
	F(\mvec{k},n,\{a_p\}) & = {n \choose a_1} \left(\prod_{p=2}^{d-1} {a_p \choose a_{p-1}}\right) \left(\prod_{p=1}^d k_p^{a_p}\right),  \label{eq:F_Taylor} \\
	\hat{Q}_{s,n}(\{a_p\},\mvec{k}) & = {i^n} F(\mvec{k}_q,n,\{a_p\}) \frac{i k_s}{|\mvec{k}_q|^2} \hat{\Theta}(\mvec{k}_q) \\
	Q_{s,n}(\{a_p\},\mvec{g}) & = \frac{1}{L^d} \sum_{q=1}^{N_q}  \hat{Q}_{s,n}(\{a_p\},\mvec{k}_q)
    \text{e}^{i \mvec{k}_q . \mvec{g}}.
\end{align}
In the above, we have used the notation ${n \choose p} = n! / (p! (n-p)!)$.
The set of indices $\{ a_p \}$ are under the constraint $\sum_{p=1} a_p=n$. The multi-dimensional array $Q_s$ is only the Fast Fourier synthesis of $\hat{Q}_{s,n}$ on a regular mesh. $\hat{Q}_{s,n}$ can be easily computed from the modes $\hat{\Theta}(\mvec{k}_q)$. Finally, $\Psi_s(\mvec{r})$ is obtained by taking all $Q_{s,n}$ and summing after multiplication by the adequate polynomial in the distance to the nearest grid point to $\mvec{r}$. 
As an example I give the expression of the cases $n=0,1$ and $2$ and $d=3$ in Table~\ref{tab:fourier_taylor_example}.

\begin{table}
	\caption{\label{tab:fourier_taylor_example} This table gives the expression of $F(\mvec{k},n,\{a_p\})$ for the first three orders and for $d=3$.} 
	\begin{center}
	\begin{tabular}{lll}
		\hline
		Order $n$ & $\{a_p \}$ & $F(\mvec{k},n,\{a_p\} )$ \\
		\hline
		$n=0$ & $a_1=a_2=a_3=0$ & 1 \\
		$n=1$ & $a_1=1$, $a_2=a_3=0$ & $k_1$ \\
		      & $a_1=0$, $a_2=1$, $a_3=0$ & $k_2$ \\
		      & $a_1=a_2=0$, $a_3=1$ & $k_3$ \\
		$n=2$ & $a_1=2$, $a_2=a_3=0$ & $k_1^2$ \\
		      & $a_1=1$, $a_2=1$, $a_3=0$ & $2 k_1 k_2$ \\
		      & $a_1=0$, $a_2=1$, $a_3=1$ & $2 k_2 k_3$ \\
		      & $a_1=1$, $a_2=0$, $a_3=1$ & $2 k_1 k_3$ \\
		      & $a_1=0$, $a_2=2$, $a_3=0$ & $k_2^2$ \\
		      & $a_1=0$, $a_2=0$, $a_3=2$ & $k_3^2$ \\
		\hline
	\end{tabular}
	\end{center}
\end{table}

I note that the number of required Fast Fourier Transforms goes quickly up as the number of elements in $\{ a_p \}$ for a given $n$. This number is exactly equal to $n+d-1 \choose n$. Fortunately, the series converges very quickly and for the problem studied in this work it is sufficient to stop at $n=3$. The time complexity of this algorithm becomes $\mathcal{O}(N_q) + \mathcal{O}(N_g \log N_g)$ instead of $\mathcal{O}(N_q N_g \log N_g)$ if the sum in Equation~\eqref{eq:psi_fourier} was carried out directly. The expected gain is huge when $N_q$ becomes high, typically for large distance surveys of several thousands of objects.

\section{Fourier-Taylor Wiener filter}
\label{app:ftaylor_wiener}

A similar problem as the one tacked in Appendix~\ref{app:interpolation} is met when one computes the mean velocity field compatible with the velocity of a set of tracers as given by Equation~\eqref{eq:c_grf_wiener}. 
The problem is the following, given a set of weights $\{ w_j\}$ and positions $\{\mvec{r}_j\}$, we need to determine the value of the modes $\hat{f}_m(\mvec{k})$:
\begin{equation}
\Hat{f}_m(\mvec{k}) =  P(|\mvec{k}|) \frac{-i k_m}{k^2} 
   \sum_{j=1}^{N_d} w_j \hat{r}_{j,m} \text{e}^{- i \mvec{k}.\mvec{r}_j} \label{eq:wiener_fourier_taylor}
\end{equation}
A priori, it is relatively simple and it can be achieved by direct summation for a time complexity of $\mathcal{O}(N_d N_q \log N_q)$ where $N_q$ is the number of modes $\mvec{k}$. It turns out that this is costly because of the big number of trigonometric functions to evaluate. Additionally, if the number of tracers becomes significant, e.g. a few thousands, the filling of this array becomes quite costly. So I propose to use another Fourier-Taylor expansion to reduce the cost.

The Equation~\eqref{eq:wiener_fourier_taylor} may be rewritten as followed:
\begin{equation}
  \Hat{f}_m(\mvec{k}) =  P(|\mvec{k}|) \frac{-i k_m}{k^2}
     \sum_{\mvec{g}} \left(\sum_{j=1}^{N_d} w_j \hat{r}_{j,m} \text{e}^{- i \mvec{k}.(\mvec{r}_j-\mvec{g})} \delta_K(\mvec{g} - \mvec{r}^{(g)}_j)\right) \text{e}^{- i \mvec{k}.\mvec{g}}, 
\end{equation}
where $\delta_K(\mvec{x}) = \prod_{j} \delta_K(x_j)$ is the Kronecker-delta $\delta_K(a)=1$ if and only if $a=0$, $\mvec{r}^{(g)}_j$ is the position of the grid point nearest to $\mvec{r}_j$.  The symbol $\sum_{\mvec{g}} {}$ means that the summation is run on all the grid points. The Kronecker-delta is well defined as we are only considering a countable set of grid point position. Now we can expand the argument of the inner exponential:
\begin{align}
 \Hat{f}_m(\mvec{k}) &=  P(|\mvec{k}|) \frac{-i k_m}{k^2}
     \sum_{\mvec{g}} \sum_{n=0}^{+\infty} \left(\sum_{j=1}^{N_d} w_j \hat{r}_{j,m} \frac{(- i)^n}{n!} \left(\mvec{k}.(\mvec{r}_j-\mvec{r}^{(g)}_j)\right)^n \delta_K(\mvec{g} - \mvec{r}^{(g)}_j)\right) \text{e}^{- i \mvec{k}.\mvec{g}} \\
      &  =  P(|\mvec{k}|) \frac{-i k_m}{k^2} \sum_{n=0}^{+\infty} \frac{(- i)^n}{n!}  \sum_{\{a_p\} } F(\mvec{k},n,\{a_p\}) \hat{G}_m(\mvec{k},n,\{a_p\})
\end{align}
with
\begin{align}
  \hat{G}_m(\mvec{k},n,\{a_p\}) & = \sum_{\mvec{g}} G_m(\mvec{g},n,\{a_p\}) \text{e}^{- i \mvec{k}.\mvec{g}} \label{eq:wiener_taylor_analysis} \\
   G_m(\mvec{g},n,\{a_p\}) & = \sum_{j=1}^{N_d} w_j \hat{r}_{j,m} \left(  \prod_{p=1}^d \left(r_j-r^{(g)}_j\right)^{a_p} \right) \delta_K(\mvec{g} - \mvec{r}^{(g)}_j),
\end{align}
with $F(\mvec{k},n,\{a_p\})$ as defined in Equation~\eqref{eq:F_Taylor}. Again we can recognize a Fast Fourier analysis step in Equation~\eqref{eq:wiener_taylor_analysis}. This analysis is run on a multi-dimensional array which has non-zero values only at grid points which are next to a tracer. The values of the field depends on moments of the distribution of tracers assigned to each grid elements. Finally, the time complexity is again now $\mathcal{O}(N_t) + \mathcal{O}(N_q \log N_q)$. 
I note that the transforms defined in this appendix have more application to compute the Fourier transform of a function sampled on a non regular grid, provided one is ready to remove the highest modes which are the most imprecise due to the Taylor expansion.

\bsp
\label{lastpage}

\end{document}